\documentstyle[preprint,tighten,prc,aps,eqsecnum,epsfig]{revtex}  
\def\be{\begin{equation}}
\def\ee{\end{equation}}
\def\bea{\begin{eqnarray}}
\def\eea{\end{eqnarray}}

\def\g5{\gamma_5}
\def\vep{\varepsilon}

\def\sl{\vec {\sigma_l}}
\def\s{\vec \sigma}
\def\ve{\vec \varepsilon}
\def\kh{\hat k}
\def\nh{\hat  \nu}
\def\lp{\lambda_+}
\def\la{\lambda}
\def\vep{\varepsilon}
\def\fot{\frac{1}{2}}
\def\Ls{\Lambda_s}
\def\Lt{\Lambda_t}
\def\tm3{\times 10^{-3}}
\def\sm1{s^{-1}}
\def\vepjnm{\varepsilon^{njm}}
\newcommand{\noi}{\noindent}

\def\Journal#1#2#3#4{{#1}{\bf #2}, #3 (#4)}

\def\NPA{{ Nucl. Phys.} {\bf A}}
\def\NPB{{ Nucl. Phys.} {\bf B}}
\def\NP{ Nucl. Phys.\,\,}
\def\PRL{ Phys. Rev. Lett.\,\,}
\def\PRC{{ Phys. Rev.} C\,\,}
\def\PRD{{ Phys. Rev.} D\,\,}
\def\PRv{ Phys. Rev.\,\,}
\def\FBS{Few--Body Syst.\,\,}
\def\PLB{{ Phys. Lett.} B\,\,}
\def\EPJC{{ Eur. Phys. J.} C\,\,}

\def\PR{ Phys. Rep.\,\,}

\def\FECAY{ Fiz. Elem. Chastits At. Yadra\,\,}
\def\SJPN{ Sov. J. Part. Nucl.\,\,}
\def\JPG{ J. Phys. G\,\,}
\def\IJMPE{{ Int. J. Mod. Phys.} E\,\,}

\begin{document}

\draft 

\title{
{\large{\bf On radiative muon capture in hydrogen}}}
\author{E.~Truhl\'{\i}k\footnote{Email address: truhlik@ujf.cas.cz}}
\address{Institute of Nuclear Physics, Academy of Sciences of the Czech Republic,
CZ--250 68 \v{R}e\v{z} n. Prague, Czechia
}

\author{F.C.~Khanna\footnote{Email address: khanna@phys.ualberta.ca}}
\address{Theoretical Physics Institute, Department of Physics, University
of Alberta, Edmonton, Alberta, Canada, T6G 2J1\,
{\em and}\, TRIUMF, 4004 Wesbrook Mall, Vancouver, BC, Canada, V6T 2A3
}

\maketitle

\begin{abstract} 
\noi
We analyze the radiative capture of the negative muon in hydrogen 
using amplitudes derived within the chiral Lagrangian approach. 
Besides the leading and next to leading order terms, given by the well-known Rood-Tolhoek Hamiltonian,
we extract from these amplitudes
the corrections of the next order in $1/M$ ($M$ is the nucleon mass). In addition, 
we estimate within the same formalism also the $\Delta(1232)$ isobar excitation effects and
processes described by an anomalous Lagrangian. Using of the 
parameters of the model obtained from the analysis of the pion photoproduction, 
which restricts the arbitrariness of the $\pi N \Delta$ and $\gamma N \Delta$
vertices off-shell, allows us to 
explain two times more of the discrepancy between the value $g^{PCAC}_P$ of the induced pseudoscalar
form factor $g_P$, predicted by partial conservation of the axial current, and of $g_P$ extracted 
from the recent TRIUMF experiment, than the standard approach.
Varying these parameters independently, one can remove the discrepancy completely.
\end{abstract}


\noi

\renewcommand{\thefootnote}{\arabic{footnote}}
\setcounter{footnote}{0}

\section{Introduction}

It is well known \cite{BS} that the charged weak interaction of the nucleon with a lepton 
is described by the weak hadron current
\be
J^a_{W,\,\mu}(q_1)\,=\,J^a_{V,\,\mu}(q_1)\,+\,J^a_{A,\,\mu}(q_1)\,,  \label{JW}
\ee
where the vector part is given by the matrix element
 of the isovector  
Lorentz 4--vector current operator between the nucleon states,
\be
\hat{J}^a_{V,\,\mu}(q_1) = i \left(g_V(q^{\,2}_1)\gamma_\mu - \frac{g_M(q^{\,2}_1)}{2M}
\sigma_{\mu\,\nu} 
q_{1\,\nu} \right)\, \frac{\tau^a}{2}\,,   \label{Vamu}
\ee
and the axial--vector part is analogously,
\be
\hat{J}^a_{A,\,\mu}(q_1) = i \left(-g_A(q^{\,2}_1)\gamma_\mu \gamma_5 + i \frac{g_P(q^{\,2}_1)}{m_l}
 q_{1\,\mu}
\gamma_5 \right)\,\frac{\tau^a}{2}\,.  \label{Aamu}
\ee
Here a is the isospin index,  $m_l$ is the lepton mass and the 4--momentum transfer
is given by $q_{1\,\mu} = p'_\mu - p_\mu$, where $p'_\mu$ ($p_\mu$) is the 4--momentum of the final 
(initial) nucleon.

The least known of the four form factors entering the currents Eqs.\,(\ref{Vamu}) and (\ref{Aamu}) 
is the induced pseudoscalar form factor, $g_P(q^{\,2}_1)$, in the axial--vector 
current $\hat{J}^a_{A,\,\mu}$. 
Actually, its presence in the axial--vector current
(\ref{Aamu}) tests our understanding of the basic strong and weak interaction 
processes, such as the strong $\pi NN$ vertex and the weak pion decay. Elementary
calculations lead to 
\be
g_P(q^{\,2}_1) = -2g_{\pi NN}f_\pi m_l\,\Delta^\pi_F(q^{\,2}_1)\,,  \label{gP}
\ee
where $\Delta^\pi_F(q^{\,2}_1)$ is the pion propagator,  $g_{\pi NN}=13.05$ is the pseudoscalar $\pi NN$ coupling
constant and  $f_\pi=92.4\,$MeV is the pion decay constant.

The matrix element of the axial current $\hat{J}^a_{A,\,\mu}$ should satisfy 
partial conservation of the axial current (PCAC).
It is easy to obtain that 
\be
\bar u(p') q_{1\,\mu} \hat{J}^a_{A,\,\mu} u(p) = \bar u(p')\left[ 2M g_A F_A(q^{\,2}_1) - 
\frac{g_P(q^{\,2}_1)}
    {m_l} q^{\,2}_1 \right]\gamma_5 \frac{\tau^a}{2} u(p)\,.   \label{divA1}
\ee
It is seen from this equation, that if 
\be
\tilde{g}_P(q^{\,2}_1)\,=\,-2g_{\pi NN}f_\pi m_l \frac{1}{q^{\,2}_1}\left[1+\frac{Mg_A}{g_{\pi NN}f_\pi}\,F_A(q^{\,2}_1)\right]
\,,  \label{gtP}
\ee
is subtracted from $g_P(q^{\,2}_1)$, then indeed, PCAC is valid.
Here we put 
\be
g_A(q^{\,2}_1) = g_A F_A(q^{\,2}_1)\,,
\ee
with $g_A \equiv g_A(0) = -1.267$. 
In the chiral model \cite{IT,STK}, the axial form factor is of the monopole form
\be
F_A(q^{\,2}_1)\,=\,\frac{m^2_{a_1}}{m^2_{a_1}+q^{\,2}_1}\,,   \label{FA}
\ee
where $m_{a_1}$ is the mass of the axial vector meson $a_1(1260)$. Then
for the $\tilde{g}_P(q^{\,2}_1)$, Eq.\,(\ref{gtP}), we have
\be
\tilde{g}_P(q^{\,2}_1)\,=\,
 -2g_{\pi NN}f_\pi m_l \frac{1}{q^{\,2}_1}\left[q^{\,2}_1+\left(1+\frac{Mg_A}{g_{\pi NN}f_\pi}\right)m^2_{a_1}\right]\,
 \Delta^{a_1}_F(q^{\,2}_1)\,.  \label{gtP1}
\ee
However, this equation for  $\tilde{g}_P(q^{\,2}_1)$ cannot be used, because of the singularity for $q^{\,2}_1=0$,
which shows its presence in the hadron radiative part of the radiative muon capture amplitude (RMC)
for large photon momentums $k$ and it is close to the physical region for the ordinary muon capture (OMC), because of
the large value of the axial meson mass. So our model cannot be used beyond the exact Goldberger--Treiman
relation and we take
\be
\tilde{g}_P(q^{\,2}_1)\,=\,
 -2g_{\pi NN}f_\pi m_l\,
 \Delta^{a_1}_F(q^{\,2}_1)\,,  \label{gtP2}
\ee
which is in agreement with \cite{AD}.
 
The best way to search for the effect of the form factor $g_P(q^{\,2}_1)$ is the muon capture.
In the elementary process of OMC in the hydrogen,
\be
\mu^-\, +\, p\, \longrightarrow\, \nu_\mu \,+\, n\,,   \label{OMCp}
\ee
according to Eq.\,(\ref{gP}), the value of the induced pseudoscalar form factor $g_P$ is
\be
g_P^{OMC}(p)\,\equiv\,g_P(q^{\,2}_1=0.877 m^2_\mu)\, =\, -\frac{2g_{\pi NN}f_\pi m_\mu}{0.877 m^2_\mu + m^2_\pi}\, = \,
6.87\,g_A\,=\,-8.71\,,  \label{gPOMCp}
\ee
and for $\tilde{g}_P(q^{\,2}_1)$, Eq.\,(\ref{gtP2}), we have
\be
\tilde{g}_P^{OMC}(p)\,\equiv\,\tilde{g}_P(q^{\,2}_1=0.877 m^2_\mu) = -\frac{2g_{\pi NN}f_\pi}{0.877 m^2_\mu 
+ m^2_{a_1}}\, = 0.13\,g_A\,=\,-0.16\,,  \label{gtPOMCp}
\ee
which is a correction of $\approx\,2\% $ to the $g^{OMC}_P(p)$, Eq.\,(\ref{gPOMCp}).
The resulting value is
\be
g_P^{PCAC}(p)=g_P^{OMC}(p)-\tilde{g}_P^{OMC}(p)=-8.55\,.    \label{gPCACp}
\ee
The axial form factor of the nucleon  has recently been measured by the $p(e,e'\pi^+)n$ reaction
in Ref.\,\cite{LAL}. The dipole form of the form factor was used and the extracted axial mass
$m_A=1.077\pm 0.039\,$GeV. For this form factor,
\be
\tilde{g}_P(q^{\,2}_1)\,=\,
 -2g_{\pi NN}f_\pi m_l\,
 \frac{2m^2_A+q^{\,2}_1}{(m^2_A+q^{\,2}_1)^2}\,.  \label{gtP3}
\ee
and analogously with Eqs.\,(\ref{gtPOMCp}) and (\ref{gPCACp}) we have
\be
\tilde{g}_P^{OMC}(p)\,=\,0.34\,g_A\,=\,-0.43\,,  \label{gtPOMCpd}
\ee
and
\be
g_P^{PCAC}(p)\,=\,-8.28\,,    \label{gPCACpd}
\ee
respectively.
Both values of $g_P^{PCAC}(p)$ are in reasonable agreement with the calculations
of $g_P^{PCAC}(p)$ within the framework of the heavy baryon chiral perturbation theory (HBChPT),
\cite{BKM,BFHM,FLMS}.

Very flat dependence of the capture rate on $g_P$ in the OMC by the proton, Eq.\,(\ref{OMCp}),
provides its world average value \cite{Bar} with an error of $\approx$ 20 \% and
particular experiments have an error larger by a factor of $\approx$ 2.

Recently, a very precise experimental study of the muon capture by $^{3}$He \cite{Vor,Ac},
\be
\mu^-\,+\,^{3}He\,\longrightarrow\,\nu_\mu\,+\,^{3}H\,,     \label{OMC3He}
\ee
yields the transition rate
\be
\Gamma_{exp}\,=\,1494\,\pm\,4\,s^{-1}\,,  \label{GHe3}
\ee
which allowed \cite{CT} an extraction of the value of $g_P$ with an 
accuracy of $\approx$ 20 \% from this experiment alone,
\be
\frac{g_P}{g_P^{OMC}(^{3}He)}\,=\,1.05\,\pm\,0.19\,,  \label{gPlgOMCHe3}
\ee
where for the reaction Eq.\,(\ref{OMC3He})
\be
g_P^{OMC}(^{3}He)\,\equiv\,g_P^{OMC}(q^{\,2}_1=0.954\,m^2_\mu)\,=\,6.68\,g_A\,.   \label{gPOMCHe3}
\ee
Analogously with Eq.\,(\ref{gPCACp}), this value of $g_P$ differs slightly from the one demanded by PCAC
by about the same amount as for the reaction Eq.\,(\ref{OMCp}).
A contribution of 20 \% due to the meson exchange current
effect turned out to be essential to get the calculated transition rate
\be
\Gamma_{th}\,=\,1502\,\pm\,32\,s^{-1}\,,
\ee
into agreement
with the data Eq.\,(\ref{GHe3}). Let us note that a further improvement of the extracted value
of $g_P$ is hindered by an uncertainty of \mbox{$\approx$ 2 \%} in calculations 
\cite{CT} which will be difficult to improve. The main uncertainty arises from 
the less known parameters of the $\Delta$ excitation processes.

Another attractive tool to extract the value of $g_P^{PCAC}$ 
is the RMC by the proton,
\be
\mu^-\, +\, p\, \longrightarrow \,\nu_\mu\, +\,\gamma\,+\, n\,.   \label{RMCp}
\ee
As is well known \cite{KlR}, the RMC amplitude contains the pseudoscalar form factor $g_P$ 
in the form
\be
g^{L(N)}_P\,=\,-x\frac{2g_{\pi NN}f_\pi m_\mu}{(q^{L(N)})^2+m^2_\pi}\,\longrightarrow
\,x\,g_P^{OMC}(p)\,\frac{0.877m^2_\mu+m^2_\pi}{(q^{L(N)})^2+m^2_\pi}\,
\left[1-\left((q^{L(N)})^2+m^2_\pi\right)\,F\left((q^{L(N)})^2\right)\right]\,, \label{gPLN}
\ee 
where we implemented the correction of Eq.\,(\ref{gtP}) and
\be
F\left((q^{L(N)})^2\right)\,=\,1/\left((q^{L(N)})^2+m^2_{a_1}\right)\,, \label{Fchm}
\ee
for the chiral model \cite{IT,STK} and
\be
F\left((q^{L(N)})^2\right)\,=\,\left((q^{L(N)})^2+2m^2_A\right)/\left((q^{L(N)})^2+m^2_A\right)^2\,,
\label{Fd}
\ee
for the dipole form of the form factor.
As it is seen from Eq.\,(\ref{gPLN}), the form factor $g_P$
depends either on the square of the four-momentum transfer $q^L\,=\,p-p'\,=\,\nu+k-\mu\,=\,-q_1$
(characterizing the muon radiation process) or on the $q^N\,=\,\nu-\mu\,=\,q^L-k$ (for the hadron
radiation). For large photon momentums $k$, $(q^L)^2\,\approx\,+m^2_\mu$, whereas $(q^N)^2\,\approx\,-m^2_\mu$, which 
enhances the hadron radiation amplitude by a factor of $\approx\,3$.  This enhancement makes the reaction (\ref{RMCp}) 
particularly interesting. On the other hand, the dependence on $g_P$ of the effective
form factors $g_i$, entering the effective RMC Hamiltonian,  appears only from the order ${\cal O}(1/M)$, 
which makes the isolation of the dependence of the photon spectrum on $g_P$ 
difficult\footnote{In Ref.\, \cite{AFM}, it is proposed to isolate the effect due to the hadron radiative amplitude
in a very difficult polarization experiment.}. The factor $x$ in Eq.\,(\ref{gPLN})
is used to study the change of the photon spectrum and capture rates by scaling $g_P$.

The theory of the RMC was elaborated by many authors during the decades 
(see Refs.\,\cite{AD},\cite{KlR,HYL,MW,LS,O,RoT,Be,ChS,F,GO,Kl,GT,BF1} and references therein).

The nuclear Hamiltonian, suitable for use in nuclear physics calculations of the RMC processes, was provided first
by Rood and Tolhoek \cite{RoT}. It contains the leading and the next to leading order terms in $1/M$ derived from
the conserved RMC amplitude given by a set of the Feynman diagrams. 
Christillin and Servadio \cite{ChS} rederived  in an elegant way the RMC amplitude
obtained earlier by Adler and Dothan \cite{AD} using the low energy theorems.
This amplitude is written in terms of elastic weak form factors and pion photoproduction amplitude, up
to terms linear in $k$ and $q$. It was also found \cite{ChS} that higher order terms cannot
be obtained using this method. Recently, this amplitude was produced \cite{STK} from 
a chiral Lagrangian of the $N \pi \rho \omega a_1$ system. 
It satisfies the corresponding continuity equations and the consistency condition exactly.
Higher order terms follow without any restriction.
It was shown that the leading order terms
coincide with those given by the low energy theorems.  However, higher order terms
differ, which is given by a different prescription to pass towards higher energies.

The above mentioned set of the relativistic Feynman diagrams was used by
Fearing \cite{F} to calculate the photon energy spectrum for the reaction Eq.\,(\ref{RMCp}). 
This work was later extended by Beder and Fearing \cite{BF1} by considering also 
the contribution from the $\Delta$ excitation processes. A recent comparison of the TRIUMF  
experiment \cite{TRIUMF1,TRIUMF2} with the Beder--Fearing calculations provided  
a value of $g^{OMC}_P(p)$ which is enhanced  by $\approx$ 50\% in comparison with the value
of Eq.\,(\ref{gPOMCp}), which corresponds to using  $g_P$ from Eq.\,(\ref{gPLN}) with  $x=1.5$ . 
This is the so--called '$g_P$ puzzle'.

In connection with the presence of the factor $x$ in Eq.\,(\ref{gPLN}), it should be noted that

(i) Referring only to the change of $g^{OMC}_P(p)$ is confusing. As it is seen from Eq.\,(\ref{gPLN}),
the whole form factor $g_P$ is scaled.

(ii) In the experiment \cite{TRIUMF1,TRIUMF2}, the high energy part of the photon spectrum is
measured. Then increasing $x$ simulates processes enhancing this part of the spectrum\footnote{ See also
the discussion in Ref.\,\cite{BHM}.}.

In our opinion, the variation of $x$ can be considered as a tool
to study an uncertainty in our knowledge of $g_P$ due to a restricted experimental accuracy.
Any real difference from $x=1$ would mean violation of PCAC.

Searching for the processes enhancing the high energy part of the photon spectrum has recently been
performed within  the concept of HBChPT by several
authors \cite{BHM,MMK,SADM,AMK,AMK1}. Ando and Min \cite{SADM} considered one--loop order 
correlations to the tree approximation and confirmed the existing discrepancy. Bernard, Hemmert and
Mei\ss ner (BHM) calculated \cite{BHM} both ordinary and radiative muon capture on the proton
in an effective field theory of pions, nucleons and $\Delta$ isobars by using the small scale expansion
\cite{HHK}. According to \cite{BHM}, the most probable explanation of the problem is a combination
of many small effects. Besides the photon spectra, BHM present the numerical results also for the 
singlet ($\Ls$) and triplet ($\Lt$) capture rates. This will enable us to compare our calculations with
those by BHM in a more detail. Here we only note the difference of $\approx$ 10\% in $\Lt$.
As we shall see later, about half of this difference arises from the use of an approximate equation for 
the neutrino energy in Ref.\,\cite{BHM}.

In Ref.\,\cite{AMK}, a possible explanation of the discrepancy was suggested that a fraction 
of spin $3/2$ orthomolecular p$\mu$p state in liquid hydrogen can exist. The analysis of the experimental
photon spectrum \cite{TRIUMF1,TRIUMF2} yielded 10 to 20\% of this state. However, this
is in sharp contrast with the existing calculations \cite{H,Ba}, which give zero fraction
of this state. As noted very recently in Ref.\,\cite{AMK1}\footnote{See also a discussion in Ref.\,\cite{AFM}.}, 
a new analysis restricts the fraction of spin $3/2$ orthomolecular p$\mu$p state to at most 5\%.

Finally, let us comment on Ref.\,\cite{CC1}, where Cheon and Cheoun reported on the derivation of 
an additional term from a chiral model, which does not appear in the standard approach to the RMC on the proton
and which generates a large contribution to the photon spectrum. As it was shown in Ref.\,\cite{ST}, 
the report \cite{CC1} suffers from two flaws. First, the derivation of this terms contains an
algebraic error due to an incorrect application of the covariant derivative in Eq.\,(14). 
After removing it, the effect is reduced by a factor of $\sim\,5$. The second flaw in Ref.\,\cite{CC1} is 
related to the introduction of the pseudovector $\pi NN$ coupling by the vertex ${\cal L}_1$ of Eq.\,(18),
which yields the desired term. However, the equivalent passage from one 
type of the $\pi NN$ coupling to another one is guaranteed only by the Foldy--Dyson unitary transformation. 
As shown in Ref.\,\cite{ST}, when this transformation is applied
to a chiral model with the pseudoscalar $\pi NN$ coupling, the pseudovector $\pi NN$ coupling appears
in the resulting Lagrangian, which does not contain the incriminating term, however. Besides, the presence of 
this term in the RMC amplitude violates the Ward--Takahashi identity derived in Ref.\,\cite{STK}.
Later attempt to improve the situation \cite{CC2} suffers from the same shortcomings.
The report \cite{CC1} was also criticised in Ref.\,\cite{HWF}.

This situation makes the expectation of the result from the next TRIUMF  experiment
in helium,
\be
\mu^-\,+\,^{3}He\,\longrightarrow\,\nu_\mu\,+\,\gamma\,+\,^{3}H\,,     \label{RMC3He}
\ee
with a particular tension. However, one should keep in mind also complications analogous
to those in reaction Eq.\,(\ref{OMC3He}) and non-negligible meson exchange current effects
are to be expected, which makes the analysis much more difficult.
It is clear that the Beder--Fearing relativistic formalism is not applicable
in calculations with the realistic 3N wave functions
and a consistent non--relativistic approach should be developed. Here we make an independent step 
in this direction by making the non--relativistic reduction of the amplitudes 
derived in Ref.\,\cite{STK} from a chiral Lagrangian of the $N\pi\rho\omega a_1$
system.
As a result, we get an effective Hamiltonian which is
close to the one obtained by Rood and Tolhoek \cite{RoT} but not identical with it.
We also apply the constructed effective Hamiltonian to compute both the capture rates and 
the photon energy spectra for the reaction Eq.\,(\ref{RMCp}) and for various
spin states. 
Our reduction provides more terms of the order ${\cal}(1/M)$ and ${\cal}(1/M^2)$ than Rood and Tolhoek present. 
Added to the leading order terms, they should reproduce with a good accuracy results given 
in Ref.\,\cite{F}\footnote{These corrections up to the order ${\cal}(1/M^2)$ were discussed in Ref.\,\cite{FS}.}. 
Another set of terms of the order ${\cal O}(1/m_\rho^2)\approx {\cal O}(1/M^2)$ is produced by reduction of
additional relativistic amplitudes following from our chiral Lagrangian. 
We shall call it hard pion (hp) correction. Numerically, it enhances the photon 
spectra by 2--4\%. 

Next we include the $\Delta$ isobar using again the formalism of chiral Lagrangians
developed in Refs.\,\cite{IT,KT}, which we extend by adopting results of 
Refs.~\cite{OO,DMW1,DMW2}. Then the resulting $N \Delta \pi\rho a_1$ Lagrangian 
consists of three terms and is characterised by three couplings and four arbitrary
parameters $A,\,X,\,Y,\,Z$. In its turn, each term contains a tensor of the form
\be
\Theta_{\mu\nu}(B)\,=\,\delta_{\mu\nu}\,+\,\left[\fot(1\,+\,4B)A\,+\,B\right] 
\gamma_\mu \gamma_\nu\,,\quad B\,=\,X,\,Y,\,Z\,,  \label{Th} 
\ee 
which ensures 
the independence of the $\Delta$ contribution to the S--matrix on the parameter 
$A$. The choice $A\,=\,-1$ simplifies the $\Delta$ propagator considerably. The 
parameters $X,\,Y,\,Z$, which reflect the off--shell ambiguity of the massive
spin 3/2 field were found \cite{OO,DMW1,DMW2,BDM,DPC} by analyzing the data on pion
photoproduction\footnote{This model describes well also the latest data on the $\pi^0$
electroproduction on the proton \cite{MITB}.}. 
The value of these parameters depends on how the pion photoproduction 
amplitude is unitarized. This model does not require the use of the Breit--Wigner form
of the $\Delta$ propagator.

In the calculations of the $\Delta$ excitation effect in the reaction Eq.\,(\ref{RMCp}), 
Beder and Fearing \cite{BF1} took a model for
needed vertices with $\Theta_{\mu\nu}=\delta_{\mu\nu}$, the Breit-Wigner form 
of the $\Delta$ propagator and the needed 
$\gamma N \Delta$ coupling from Ref.\,\cite{DMW1}, thus introducing 
an inconsistency into calculations. This model provides about 7\% effect from
the $\Delta$ excitation to the photon spectrum.

The $\pi N \Delta$ and $\gamma N \Delta$ vertices including the off--shell parameters $X$, $Y$ and $Z$
were discussed in Ref.\,\cite{BKMr} and
the $\pi N \Delta$ vertex of the form Eq.\,(\ref{Th}) was considered also in the small scale 
expansion \cite{HHK}.
However, the dependence of the results on these parameters was not exploited in any of
the calculations performed within the framework of HBChPT.

Let us note that besides the adopted model \cite{OO,DMW1,DMW2}, other models
\cite{MAID,SL} were developed to describe the production of pions on protons
by the electromagnetic interaction. All these models consider the same non--resonant
Lagrangian of the $N \pi \rho \omega$ system, but differ principally in the treatment
of the $\Delta$ isobar and in the method of unitarization of the $\pi N$ amplitude.

We have also analyzed the contribution due to amplitudes constructed from an anomalous
Lagrangian of the $\pi \rho \omega a_1$ system \cite{STK1,TSK}. We have found that 
the influence of this contribution on the photon energy spectrum is not significant.
Earlier estimate of a contribution which arises from the Wess--Zumino--Witten
part of the anomalous Lagrangian was reported in Ref.\,\cite{FLMS1}.

One can find in the literature an attempt to study the form factor $g_P$ 
in the reaction of
electroproduction of charged soft pions off the proton \cite{BHM,CH},
\be
e\,+\,p\,\longrightarrow\,e'\,+\,\pi^+\,+\,n\,.   \label{CPP}
\ee
The starting point
of this attempt is the soft pion production amplitude given as \cite{ADa}
\be
f_\pi\,M^{nj}_\lambda(q,k)\,\stackrel{q\,\rightarrow\,0}{\longrightarrow}\,i q_\mu\,
\left<p\,'|\int d^4 y e^{-iqy}\,T\left(\hat{J}^n_{A,\,\mu}(y)\,\hat{J}^j_{V,\,\lambda}(0)\right)|p\right>
+\vepjnm\left<p\,'|\hat{J}^m_{A,\,\lambda}(0)|p\right>\,.   \label{spMnjl}
\ee
The matrix element of the time--ordered product of the two currents is related
to the RMC amplitude by the time reversal. The form factor $g_P$ is contained on
the right--hand side of Eq.\,(\ref{CPP}) in the
matrix element of the axial current. If one admits that in the soft pion limit only the 
nucleon Born terms contribute to the divergence of the current--current amplitudes,
then one has the pion production amplitude which can provide information on  $g_P$.
However, when one of the currents is axial, a contribution to the divergence of the
current--current amplitudes from the pion pole term in the t--channel survives even in the
soft pion limit \cite{DR,ITr,DT,T}. A part of this contribution cancels the induced pseudoscalar term in
the axial current and the remaining part is just the pion pole production amplitude, as
one can expect intuitively. Then in the soft pion regime, the reaction Eq.\,(\ref{CPP}) is suitable to study 
the weak axial nucleon form factor $F_A(k^2)$ and the electromagnetic form factor 
(the electromagnetic radius) of the charged pion,
but not to extract any information on $g_P$.

In our opinion, the RMC reactions and particularly reactions (\ref{RMCp}) and (\ref{RMC3He}) are at 
present the only available tool to study the form factor $g_P$ as a function of the  momentum
transfer.

In order to compare our effective form factors with the results of Ref.\,\cite{RoT}, 
we define in Sect.\,\ref{CH1} the effective Hamiltonian analogously and we consider
the velocity independent part only. 
Then in Sect.\,\ref{CH2}, we present the results for the form factors following 
from our amplitudes \cite{STK} up to the order ${\cal O}(1/M^2)$.
Further we deal with the contribution to $g_i$'s from the $\Delta$ excitation amplitudes of
our model and we compare our effective weak $N\Delta$ vertex with the one used in 
Ref.\,\cite{BF1}. 
Finally, we discuss the RMC amplitudes stemming from the anomalous Lagrangian.

In Sect.\,\ref{CH3}, we give the numerical results for the capture rates and we present various photon spectra. 
Our conclusions are presented in Sect.\,\ref{CH4}.
Our main result is an enhancement up to \mbox{$\approx$ 7--15 \%} relative to the calculations
without including the $\Delta$ isobar, which can explain twice as much of the discrepancy
in comparison with the standard approach 
between the $g_P^{PCAC}$ and
$g_P$ extracted from the experiment \cite{TRIUMF1,TRIUMF2},
if we take the values of the parameters of the model found from 
the data on pion photoproduction \cite{DMW1,DMW2,BDM,DPC}. If we 
are allowed to change these parameters independently, we can generate 
for the value of $g_P\,\approx\,g^{PCAC}_P$
the photon spectrum, which is in the region of the photon energies $k\ge$ 60 MeV close to the spectrum, 
that should correspond to the experimental one.

Our triplet capture rate agrees with the one calculated very recently by BHM up to 10\%. A half of this
discrepancy can be attributed to an incorrect integration over the phase volume in Ref.\,\cite{BHM}.

\section{Effective Hamiltonian for RMC}
\label{CH1}

In presenting the effective Hamiltonian, we follow Rood and Tolhoek \cite{RoT}. Then the velocity
independent part is
\bea
H^{\,(0)}_{\,eff}\,&=\,&\frac{1}{\sqrt{2}m_\mu}\,(1-\sl \cdot \nh )
\left[g_1\left(\sl \cdot \ve \right) + g_2 \left(\s \cdot \ve \right) + 
g_3 i \left(\s \cdot \ve \times \sl \right) \right. \nonumber \\
& & \left. + g'_4 \left(\sl \cdot \ve \right) \left(\s \cdot \kh \right) +
g''_4 \left(\sl \cdot \ve \right) \left(\s \cdot \nh\right) + 
g'_5 \left(\sl \cdot \kh\right) \left(\ve \cdot \nh \right) +
g'_6 \left(\ve \cdot \nh\right) \right. \nonumber \\
& & \left. + g'_7 i \left(\s \cdot \kh \times \ve\right)
+ g''_7 i \left(\s \cdot \nh \times \ve\right)
+ g'_8 \left(\sl \cdot \kh\right)\left(\s \cdot \ve \right)
+ g''_8 \left(\sl \cdot \nh\right)\left(\s \cdot \ve \right) \right. \nonumber \\
& & \left. + g'_9 \left(\sl \cdot \s\right) 
+ g'_{10}\left(\s \cdot \kh\right) \left(\ve \cdot \nh \right)
+ g''_{10}\left(\s \cdot \nh\right) \left(\ve \cdot \nh \right) \right. \nonumber \\
& & \left. +g'_{11}\left(\sl \cdot \kh\right)\left(\s \cdot \kh\right) 
\left(\ve \cdot \nh \right)
+g''_{11}\left(\sl \cdot \kh\right)\left(\s \cdot \nh\right) 
\left(\ve \cdot \nh \right) \right]\,.  \label{H0eff}
\eea

Here $\vec {\sigma}_l$ ($\vec \sigma$) are the lepton (nucleon) spin Pauli matrices and
$\hat \nu$ ($\hat k$) is the unit vector in the direction of the neutrino (photon) momentum vector
$\vec \nu$ ($\vec k$). Not all the form factors are independent. Using
equation
\be 
\ve_\lambda\,=\,-i\lambda\left(\kh\times\ve_\lambda\right)\,,\qquad 
\ve_\lambda\,=\,\frac{1}{\sqrt{2}}\left(\hat i\,-\,\lambda\hat j\right)\,, \label{vel}
\ee
one gets redefinitions:
\be
g_2\,\rightarrow\,g_2-\lambda \left(g'_7+y g''_7 \right)\,,\quad
g'_{10}\,\rightarrow\,g'_{10}+\lambda g''_7\,,\quad
g'_8\,\rightarrow\,g'_8+\lambda g_3\,,\quad
g'_4\,\rightarrow\,g'_4-\lambda g_3\,,  \label{redgis}
\ee
where $y=(\nh \cdot \kh)$. The last two terms in (\ref{H0eff}) are new in comparison with \cite{RoT}.

\section{Contribution to $H^{\,(0)}_{eff}$ from the amplitudes of the chiral Lagrangian
of the $N\Delta \pi \rho \omega a_1$ system}
\label{CH2}

Here we  discuss our amplitudes and contributions to $g_i$'s.
We start by presenting briefly the part of the RMC amplitude derived earlier \cite{STK} without
$\Delta$'s,
referring for details to Sect.\,3 of that paper. Then we deal
with the amplitudes describing the $\Delta$ excitation processes and we compare our 
effective vertices with those of Ref.\,\cite{BF1}. Finally, we discuss the amplitudes
stemming from the anomalous Lagrangian.

\subsection{The RMC amplitude without $\Delta$'s}
\label{CH2.1}

Besides the muon
radiative part $M^a(k,q)$, the amplitude $T^a(k,q)$ \cite{STK} consists of three terms representing
the hadron radiative amplitude 
\begin{equation}
T^a(k,q)  =  \frac{eG}{\sqrt{2}}\,\left\{\,M^a(k,q)\,+\,l_\mu(0)\epsilon_\nu(k)
\,\left[\,M^{B,\,a}_{\mu \nu}(k,q)\,+\,M^a_{\mu \nu}(\pi;k,q)\,+\,
M^{a}_{\mu \nu}(a_1;k,q)\,\right]\,\right\}\,, \label{Ta}
\end{equation}
The amplitude $M^{B,\,a}_{\mu \nu}(k,q)$ consists of the nucleon Born terms 
and of some related contact amplitudes.
The amplitude $M^a_{\mu \nu}(\pi;k,q)$ contains the mesonic amplitude $M^{m.\,c.\,,a}_{\mu \nu}(\pi;k,q)$
and all contact terms where the electroweak vertex is connected with the nucleon by the pion line.
The amplitude $M^a_{\mu \nu}(a_1;k,q)$ has graphically a similar structure as the amplitude 
$M^a_{\mu \nu}(\pi;k,q)$ with the pion line changed for the $a_1$ meson one. These amplitudes satisfy separately continuity
equations when contracted with the four momentum transfer $q_\mu$ of the weak vertex.

Since our model respect vector dominance and PCAC,
the sum of the hadron radiative amplitudes satisfies exactly the following Ward--Takahashi identities
\begin{equation}
q_\mu\,[M^{B,\,a}_{\mu \nu}\,+\,M^a_{\mu \nu}(\pi)\,+\,
M^{c.\,t.,\,a}_{\mu \nu}(a_1)]\,=\,i f_\pi m^2_\pi
\Delta^\pi_F(q)\,{\cal M}^a_{\pi,\,\nu}\,+\,
i\varepsilon^{3\,a\,b} \bar u(p')\,\hat J^b_{W,\,\nu}(q_1)\,u(p)
\,, \label{PCAC}
\end{equation}
\begin{equation}
k_\nu\,[M^{B,\,a}_{\mu \nu}\,+\,M^a_{\mu \nu}(\pi)\,+\,
M^{c.\,t.,\,a}_{\mu \nu}(a_1)]\,=\,
i\varepsilon^{3\,a\,b} \bar u(p')\,\hat J^b_{W,\,\mu}(q_1)\,u(p)
\,. \label{CVC}
\end{equation}
Besides, the monopole electroweak form factors with $m_V=m_\rho$ and $m_A=m_{a_1}$ appear
naturally in our amplitudes.
Let us note that Eq.\,(\ref{CVC}) guarantees the gauge invariance of the model.
The consistency condition \cite{ChS} for our amplitudes is
\be
\Delta^\pi_F(q)\,k_\nu{\cal M}^a_{\pi,\,\nu}\,=\,\Delta^\pi_F(q_1)\,i\varepsilon^{3\,a\,b}
M^b_\pi\,.   \label{CC}
\ee
Here ${\cal M}^a_{\pi,\,\nu}$ is the radiative pion absorption amplitude, 
$M^b_\pi$ is the pseudoscalar $\pi NN$ vertex and $q=k+q_1$.  

The leading amplitudes are the nucleon Born terms
$M^{B,\,a}_{\mu\nu}(k,q)$\footnote{For notations see Sect.\,3 of Ref.\,\cite{STK}.}
(corresponding to $M(b),\,M(c),\,M(d)$ in Ref.\,\cite{RoT}),
the amplitudes $M^a_{\mu\nu}(\pi,1)$, $M^a_{\mu\nu}(\pi,2)$ and
$M^{B,\,a}_{\mu\nu}(5)$ (the sum of them corresponds to $M(e)$ in Ref.\,\cite{RoT}),
and the mesonic amplitude $M^{m.\,c.,\,a}_{\mu\nu}$ (in correspondence with $M(f)$
in Ref.\,\cite{RoT}). As discussed above, besides these amplitudes, other contact terms appear.

The low energy theorems allow one \cite{AD,ChS}, by applying current conservation and PCAC
to a general amplitude, to determine consistently the amplitude for the RMC in terms of
elastic weak form factors and pion photoabsorption amplitude, up to terms linear
in k and q. As shown in Ref.\,\cite{ChS}, higher order terms cannot be predicted. Since our amplitudes
satisfy exactly current conservation and PCAC, we can obtain terms of any desired order.
We now present the expansion of our non--resonant amplitudes up to the order ${\cal O}(1/M^2)$.

\subsubsection{Corrections up to the order ${\cal O}(1/M)$}
\label{CH2.1.1}

The non-relativistic reduction of these amplitudes yields the following contributions 
up to the order ${\cal O}(1/M)$ to the form factors $g_i$
\bea
g_1&\,=\,&-\lp g^L_V\left[1+\frac{\vec s}{2M}\cdot\kh\right]-g^N_V\eta
+g^N_A\la\eta\mu_V\,, \nonumber  \\
g_2&\,=\,&-\lp g^L_A -g^N_P\eta + g^N_V\la\eta\mu_V - g^N_A\eta\,, \nonumber  \\
g_3&\,=\,&-\lp g^L_A +g^N_M\eta + g^N_V\eta - g^N_A\la\eta\mu_S\,, \nonumber  \\
g_4&\,=\,&\lp g^L_A - g^N_V\la\eta\mu_V + g^L_P\left[\lambda_- + \frac{\nu}{m_\mu}
(1-y)\lp\right]\,,\quad g'_4=g_4\frac{k}{2M}\,,\quad g''_4=g_4\frac{\nu}{2M}\,,
\nonumber \\
g_5&\,=\,&\lp g^L_V +g^N_M\eta\mu_V + g^N_V\eta\mu_V + g^N_M\la\eta\,,\quad 
g'_5=g_5\frac{\nu}{2M}\,,  \nonumber \\
g_6&\,=\,&\lp g^L_V + (g^N_P+g^N_A)\la\eta\mu_V + g^N_M\eta - g^N_V\eta(1+2\mu_n)\,,\quad 
g'_6=g_6\frac{\nu}{2M}\,,  \nonumber \\
g_7&\,=\,&\lp (g^L_V + g^L_M) + (g^N_P+g^N_A)\la\eta\mu_V - g^N_V\eta\,,\quad
g'_7=g_7\frac{k}{2M}\,,\quad g''_7=g_7\frac{\nu}{2M}\,,  \nonumber \\
g_8&\,=\,&-\lp (g^L_V + g^L_M) - g^N_V\la\eta\mu_S\,,
g'_8=g_8\frac{k}{2M}\,,\quad g''_8=g_8\frac{\nu}{2M}\,,  \nonumber \\
g_9&\,=\,&\lp (g^L_V + g^L_M) + (g^N_V+g^N_M)\la\eta\mu_S - 2g^N_A\eta\mu_n\,,\quad
g'_9=g_6\frac{\nu}{2M}\,,  \nonumber \\
g_{10}&\,=\,&g^N_P\eta\frac{4M\nu}{m^2_\pi+(q^L)^2} + \lp g^L_P\frac{\nu}{m_\mu}\,,\quad
g'_{10}=g_{10}\frac{k}{2M}\,,\quad g''_{10}=g_{10}\frac{\nu}{2M}\,,  \nonumber \\
g_{11}&\,=\,&\lp g^L_P\frac{\nu}{m_\mu}\,,\quad
g'_{11}=g_{11}\frac{k}{2M}\,,\quad g''_{11}=g_{11}\frac{\nu}{2M}\,.  \label{gis0}
\eea
Here our notations mostly follow Ref.\,\cite{RoT}:
\begin{displaymath}
\vec s\,=\,\vec k + \vec \nu\,,\quad 
\eta\,=\,\frac{m_\mu}{2M}\,,\quad
\la_\pm\,=\,\frac{1}{2}(1\pm \la)\,.  
\end{displaymath}
In addition we have
\be
\mu_V\,=\,1+\mu_p-\mu_n\,\equiv\,1+\kappa_V,\quad \mu_S\,=\,1+\mu_p+\mu_S\,
\equiv\,1+\kappa_S.  
\ee
Besides the  obvious momentum dependence of the form factors $g^{L(P)}_P$ given in Eq.\,(\ref{gPLN}),
all other nucleon weak vector 
and axial--vector form factors are assumed to have either the monopole momentum dependence, which
naturally appears in our model, with $m_V=m_\rho$ and $m_A=m_{a_1}$, or, for the sake of comparison,
the dipole one with $m_V=0.843\,$GeV and $m_A=1.077\,$GeV \cite{LAL}.

\subsubsection{Corrections of the order ${\cal O}(1/M^2)$}
\label{CH2.1.2}

Here we have two groups of contributions. The first one arises from the expansion of the 
amplitudes considered above by one order more in $1/M$ which leads to
\bea
\left(\frac{2M}{\eta}\right)\,\Delta g_1 &\,=\,& g^N_M\left(k - \mu_V\,\vec{\nu}\cdot\kh\right)
 - \left(g^N_V \mu_S - g^N_A
\la \eta \mu_n \right)\,\left(\vec{s} \cdot \kh\right)\,,           \nonumber \\
\left(\frac{2M}{\eta}\right)\,\Delta g_2 &\,=\,& -g^N_M \la \left(\vec{\nu} \cdot \kh \right)
+ 2 \mu_n\,\left(g^N_A + \la g^N_V - g^N_P\right) \,\left(\vec{s} \cdot \kh\right)\,, \nonumber \\
\left(\frac{2M}{\eta}\right)\,\Delta g_3 &\,=\,& -g^N_M k + 2 g^N_A \la \mu_n 
\,\left(\vec{s} \cdot \kh\right) - 2 g^N_V \mu_n\,\left(\vec{\nu} \cdot \kh\right)\,,
\nonumber  \\
\left(\frac{2M}{\nu\eta}\right)\,\Delta g''_4 &\,=\,& -g^N_M \la  \mu_S  - 2 g^N_V
\la  \mu_n\,, \nonumber \\
\left(\frac{M}{\nu\eta}\right)\,\Delta g''_8 &\,=\,& g^N_V \la \mu_n\,, \nonumber \\
\left(\frac{M}{\nu\eta}\right)\,\Delta g'_{10} &\,=\,& \left(g^N_P-g^N_A\right)\mu_n\,. 
\label{dgis1}
\eea
The main part of the contribution to the photon spectrum arises from the terms proportional to
$g^N_A$ and $g^N_V$ in $\Delta g_2$. These terms appear due to the neutron recoil induced by 
the time component of the 
weak current. Actually, the terms $\Delta g''_4$, $\Delta g''_4 $ and $g'_{10}$ contribute in the order
${\cal O}(1/M^3)$. We have verified that they change the singlet capture rate by $\approx\,10\% $ and
the triplet capture rate by $\approx\,0.8\% $, which is the reason to keep them. They also arise
presumably from the neutron recoil.

The second group of corrections to order ${\cal O}(1/m^2_\rho)\,\approx\,{\cal O}(1/M^2)$
(the hp correction) stems from some contact terms
present in the hadron radiative part of the
amplitude Eq.\,(\ref{Ta}). It is discussed in Sect.\,4 of Ref.\,\cite{STK}.
Here we quote the results of the non--relativistic reduction
\be
\Delta g_1 \,=\,-2 g^L_V \left(\frac{2M}{m_\rho}\right)^2 \eta \frac{k}{2M}\,,\quad
\Delta g_2 \,=\,-\frac{g^N_A}{2} \left(\frac{2M}{m_\rho}\right)^2 \eta \frac{2k+y\nu}{2M}\,.
 \label{dgis2}
\ee

\subsection{The RMC amplitude with $\Delta$'s}
\label{CH2.2}

We derive the RMC amplitudes arising due to the $\Delta$ excitations from chiral Lagrangians
\cite{KT,OO,DMW1,DMW2}. 
They correspond to the standard nucleon Born terms with the $\Delta$ isobar
instead of nucleon in the intermediate state. The needed Lagrangian reads
\be
{\cal L}^{M}_{\,N\Delta\pi\rho a_{1}}  = \frac{f_{\pi N \Delta}}
{m_{\pi}}\,\bar{\Psi}_{\mu}\vec{T}{\cal O}_{\mu\nu}(Z)\Psi\cdot
\left(\partial_{\nu}\vec{\pi}
+2 f_{\pi}g_\rho\vec{a}_{\nu}\right)
-g_{\rho}\frac{G_{1}}{M}\bar{\Psi}_{\mu}\vec{T}{\cal O}_{\mu\eta}(Y)
\gamma_{5}\gamma_{\nu}\Psi\cdot
\vec{\rho}_{\eta\nu}\,+\,h.\,c.   \label{LNDPRA1}
\ee
Here $\vec{T}$ is the operator of the transition spin. Another possible term in 
the $\rho N \Delta$ vertex
is suppressed by one order in $1/M$ and it does not contribute in any sizeable manner \cite{BF1}.

We take the operator ${\cal O}_{\mu\nu}(B)$ in a form \cite{OO,DMW1,DMW2}
\bea
{\cal O}_{\mu\nu}(B)\,&=&\,\delta_{\mu\nu}\,+\,C(B)\,\gamma_\mu\,\gamma_\nu\,, \label{OmnB} \\
C(B)\,&=&\,\frac{1}{2}\left(1\,+\,4B\right)\,A\,+\,B\,.  \label{CeB}
\eea
A choice $A=-1$ simplifies considerably  \cite{OO} the propagator of the $\Delta$.

The coupling constant
$f_{\pi N \Delta}$ is not well known and the values
for $f^2_{\pi N \Delta}/4\pi$
from the interval
between 0.23 and 0.36  can be found in the literature \cite{CT}.
From the dispersion theory \cite{TF},
$f^2_{\pi N \Delta}/4\pi \,\approx \,0.30$
  and
$f^2_{\pi N \Delta}/4\pi \,\approx \,0.35$ from the decay width \cite{SvH}.  Also a good fit to
the 33 phase shift was obtained in Refs.\,\cite{DMW1,DMW2}  by using 
$f^2_{\pi N \Delta}/4\pi \,\approx \,0.314$. The new data on pion photoproduction prefer 
$f^2_{\pi N \Delta}/4\pi \,\approx \,0.371$ \cite{DPC}. The ranges of the other relevant parameters of the model
are \cite{DMW2,BDM,DPC}
\be
-0.8\,\le\,Z\,\le\,0.7\,,\quad\,-1.25\,\le\,Y\,\le\,1.75\,,\quad\,1.97\,\le\,G_1\,\le\,2.65\,.  \label{ZYG1}
\ee

Our radiative amplitude with the $\Delta$ excitation can be written analogously with the
nucleon Born term $M^{B,\,a}_{\mu\nu}(1)$ \cite{STK} as
\bea
\hspace{-2cm}
M^{\Delta,\,a}_{\mu \nu} & = & -{\bar u}(p')\left[\,\left({\hat J}_{W,\,\mu\alpha}(-q)\right)^+\,
S^{\alpha \gamma}_F(Q)\hat J_{em,\,\nu \gamma}(k)\,\left(T^+\right)^a\,T^3
\right . \nonumber \\
& & \hspace{2cm} \left . +\,\left({\hat J}_{em,\,\nu\gamma}(-k)\right)^+\,S^{\gamma\alpha}_F(P)
\hat J_{W,\,\mu\alpha}(q)\,\left(T^+\right)^3\,T^a\right]u(p)\,,  \label{RMCAD} 
\eea
Here the weak $N\Delta$ vertex reads
\be
\hspace{-1cm}
{\hat J}_{W,\,\mu\alpha}(q) = {\hat J}_{V,\,\mu \alpha}(q) - {\hat J}_{A,\,\mu \alpha}(q)\,, \label{JWD}
\ee
with the vector part defined as
\be
{\hat J}_{V,\,\mu \alpha}(q) = i\left(\frac{G_1}{M}\right)m^2_\rho \Delta^\rho_F(q)
\left(q_\beta \delta_{\mu\lambda} - q_\lambda \delta_{\mu\beta}\right)\,
{\cal O}_{\alpha\beta}(Y)\,\gamma_5 \gamma_\lambda\,, \label{JWVD}
\ee
and with the axial--vector part of the form
\be
{\hat J}_{A,\,\mu \alpha}(q) = 
\left(\frac{f_\pi f_{\pi N \Delta}}{m_\pi}\right)\left[ m^2_{a_1}
\Delta^{a_1}_{\mu\lambda}(q)
-q_\mu q_\lambda \Delta^\pi_F(q)\right]\,{\cal O}_{\alpha\lambda}(Z)\,.  \label{JWAD}
\ee
Further the electromagnetic  $\gamma N\Delta$ vertex is
\be
{\hat J}_{em,\,\nu \gamma}(k) = -{\hat J}_{V,\,\nu \gamma}(k,\,k^2=0) = 
-i\left(\frac{G_1}{M}\right)\left(k_\beta 
\delta_{\nu\lambda} - k_\lambda \delta_{\nu\beta}\right)\,
{\cal O}_{\gamma\beta}(Y)\,\gamma_5 \gamma_\lambda\,. \label{JEMD}
\ee
Finally, $S^{\alpha \gamma}_F(p)$ is the $\Delta$ isobar propagator. 
With the choice ${\cal O}_{\alpha\beta}=\delta_{\alpha\beta}$, our amplitudes 
Eqs.\,(\ref{JWD})--(\ref{JWAD}) coincide in the form with the ones obtained in Ref.\,\cite{BF1} from
the study of the weak $N$--$\Delta$ vertex in the reaction 
\mbox{$\nu\,d\,\rightarrow\,\mu^-\,\Delta^{++}\,n$}\,\footnote{For a recent study of this reaction
see \cite{ASV}.}.
 
For the divergence of the resonant amplitude $M^{\Delta,\,a}_{\mu \nu}$ from Eq.\,(\ref{RMCAD}) we have
\be
q_\mu M^{\Delta,\,a}_{\mu \nu}\,=\,i f_\pi m^2_\pi \Delta^\pi_F(q^2)\,M^{\Delta,\,a}_{\pi,\,\nu}. \label{divMD}
\ee
Here the associated resonant radiative pion absorption amplitude $M^{\Delta,\,a}_{\pi,\,\nu}$ is,
\bea
M^{\Delta,\,a}_{\pi,\,\nu} & = & -{\bar u}(p')\left[\, \left( {\hat M}^{\Delta}_{\pi,\,\alpha}(-q)\right)^+\,
S^{\alpha \gamma}_F(Q)\hat J_{em,\,\nu \gamma}(k)\,\left(T^+\right)^a\,T^3
\right . \nonumber \\
& & \hspace{2cm} \left . +\,\left(\hat J_{em,\,\nu\gamma}(-k)\right)^+\, S^{\gamma\alpha}_F(P)
 {\hat M}^{\Delta}_{\pi,\,\alpha}(q)\,\left(T^+\right)^3\,T^a\right]u(p)\,,  \label{pMCAD} 
\eea
and the $\pi N \Delta$ vertex reads
\be
{\hat M}^{\Delta}_{\pi,\,\alpha}(q) = i\frac{f_{\pi N \Delta}}{m_\pi}q_\lambda
\,{\cal O}_{\alpha\lambda}(Z)\,.  \label{MpND} 
\ee

We now present the contributions from the amplitude Eq.\,(\ref{RMCAD}) to the form factors $g_i$.
They are
\bea
\Delta g_1&\,=\,&\frac{2}{3}\la(C_- - C_+)C\left\{1+(1-R)\left[C(Y)+C(Z)+2(2+R)C(Y)C(Z)
\right]\right\}\,,  \nonumber \\
\Delta g_2&\,=\,&\frac{1}{3}(C^+ + C_-)C(1-R)\left[-(1+2R)+2(1-2R)C(Y)+2(1-R)C(Z)
\right. \nonumber  \\
&& \left. \hspace{-3mm}+4(2-R)C(Y)C(Z)\right]  
\,+\,(C^+ + C_-)C\frac{g^N_P}{6Mg^N_A}\left<-\left[(1-R)(1+2R)k
\right.\right.  \nonumber \\
&& \left.\left. \hspace{-3mm}+y \nu \right]
+2(1-R)\left\{\left[(1-2R)k+y \nu\right]C(Y) + \left[(1-R)k+y \nu\right]C(Z) 
\right.\right. \nonumber \\
&& \left.\left.\hspace{-3mm}+2\left[(2-R)k+(2+R)y \nu\right]C(Y)C(Z)\right\}
\right>\,, \nonumber \\
\Delta g_3&\,=\,&\frac{2}{3}\la(C_+ + C_-)C\left\{1+(1-R)\left[C(Y)+C(Z)+2(2+R)C(Y)C(Z)
\right]\right\}\,,  \nonumber \\
\Delta g'_4&\,=\,&-\Delta g'_8\,=\,(C_+ + C_-)C\,, \nonumber \\  
\Delta g_6&=&-\la \frac{\nu}{3M}\frac{g^N_P}{g^N_A}(C_+ - C_-)C\left\{
1+(1-R)\left[C(Y)+C(Z)+2(2+R)C(Y)C(Z)\right]\right\}\,,  \nonumber  \\
\Delta g'_{10}&=&(C_+ + C_-)C\frac{\nu}{6M}\frac{g^N_P}{g^N_A}\left\{
1-2(1-R)\left[C(Y)+C(Z)+2(2+R)C(Y)C(Z)\right]\right\}\,, \label{dgisd}
\eea
where 
\bea
C&\,=\,&-\frac{4}{3}\frac{f_\pi f_{\pi N \Delta}}{m_\pi}G_1 \eta k\,,
\quad R\,=\,M/M_\Delta\,,   \nonumber \\
C^{-1}_+&\,=\,&\left\{(M_\Delta-M)+\frac{2M}{M_\Delta+M}\left[m_\mu-\nu+\frac{m_\mu}{2M}
\left(2\nu-m_\mu-\frac{2\nu}{m_\mu}(\nu+yk)\right)\right]\right\}\,, \nonumber  \\
C^{-1}_-&\,=\,&\left\{(M_\Delta-M)+\frac{2M}{M_\Delta+M}\left[\nu-m_\mu+\frac{m_\mu}{2M}
\left(2\nu-m_\mu\right)\right]\right\}\,.   \label{Cisd}
\eea

According to the concept developed in Refs.\,\cite{OO,DMW1,DMW2}, we take the mass of 
the $\Delta$ isobar real. 

\subsection{The RMC amplitude from an anomalous Lagrangian of the $\pi \rho \omega a_1$ system}
\label{CH2.3}

\begin{figure}[h!]
\centerline{
\epsfig{file=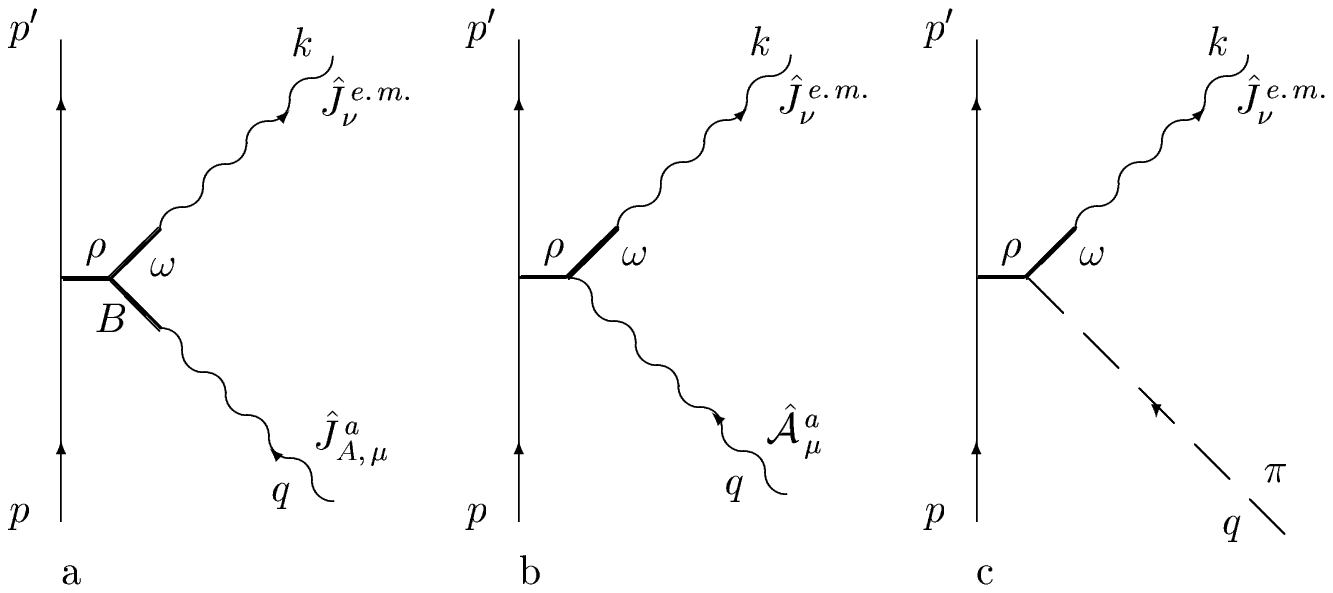}
}
\vskip 0.4cm
\caption[99]{(a),(b) -- 
the radiative hadron amplitudes obtained from the anomalous Lagrangian 
of the $\pi \rho \omega a_1$ system, Eq.\,(\ref{Lan});
in (a), B=$\pi$ or $a_1$; (c) -- the associated radiative pion absorption amplitude.
These amplitudes satisfy PCAC.
}
\label{figRMCAA}
\end{figure}

We have considered so far the amplitudes where a natural parity does not change in any vertex.
The natural parity of a particle is defined as $P\,(-1)^J$, where $P$ is the intrinsic parity and 
$J$ is the spin of the particle. Some amplitudes of this kind relevant for the process
under study are presented in Fig.\,\ref{figRMCAA}.
\noindent
The starting point is an anomalous Lagrangian of the $\pi \rho \omega a_1$ system \cite{TSK}, \cite{STK1}
constructed within the approach of hidden local symmetries \cite{BKY,M}. The 
electromagnetic interaction in such a system was first considered in Ref.\,\cite{KM} and the
relevant constants $\tilde c_i$ were extracted from the data as well. The weak
interaction was incorporated explicitly in Ref.\,\cite{STK1} and the refit of the constants to
the modern data \cite{PDG} was made in Ref.\,\cite{TSK}.

The Lagrangian reads
\bea
{\cal L}_{an} &\,=\,&2ig_\rho\vep_{\kappa \lambda \mu \nu}\,\left[
\left(\partial_\kappa \omega_\lambda\right)\,\left( g_\rho \vec {\rho}_\mu\,
-\,e\vec {\cal V}_\mu\right)
+\,\left(g_\rho\omega_\kappa\,-\,\frac{1}{3}e\,{\cal B}_\kappa\right)
\left(\partial_\lambda \vec {\rho}_\mu\right)\right] \nonumber \\
&&\cdot\,\left[{\tilde c}_7\left(\frac{1}{f_\pi}\,\partial_\nu \vec {\pi}\,
+\,e\vec {\cal A}_\nu\right)\,+\,{\tilde c}_8 \left(\fot e\vec {\cal A}_\nu)
-g_\rho \vec {a}_\nu \right)\right] \nonumber \\
&&\,+2ie\vep_{\kappa \lambda \mu \nu}\,\left[
\left(\frac{1}{3}\partial_\kappa {\cal B}_\lambda\right)\,\left( g_\rho \vec {\rho}_\mu\,
-\,e\vec {\cal V}_\mu\right)
+\,\left(g_\rho\omega_\kappa\,-\,\frac{1}{3}e\,{\cal B}_\kappa\right)
\left(\partial_\lambda \vec {\cal V}_\mu\right)\right] \nonumber \\
&&\cdot\,\left[{\tilde c}_9\left(\frac{1}{f_\pi}\,\partial_\nu \vec {\pi}\,
+\,e\vec {\cal A}_\nu\right)\,+\,{\tilde c}_{10} \left(\fot e\vec {\cal A}_\nu)
-g_\rho \vec {a}_\nu \right)\right]\,, \label{Lan}
\eea
where besides the meson fields,
the external vector isoscalar ${\cal B}_\mu$ 
and isovector $\vec {\cal V}_\mu$
and  axial vector isovector $\vec {\cal A}_\mu$ fields are also included. The constants
${\tilde c}_i$ are \cite{TSK}
\be
{\tilde c}_7=8.64\times 10^{-3}\,,\,\,{\tilde c}_8=-1.02\times 10^{-1}\,,\,\,
{\tilde c}_9=9.23\times 10^{-3}\,,\,\,{\tilde c}_{10}=1.29\times 10^{-1}\,.  \label{cti}
\ee

The axial RMC amplitudes arising from the anomalous Lagrangian ${\cal L}_{an}$, Eq.\,(\ref{Lan}), are
\bea 
M^{an,\,a}_{\mu\nu}(1)&\,=\,&i\frac{g^2_\rho}{3}\,
\vep_{\eta \nu \beta \sigma}\,k_\eta q_\sigma q_\mu \Delta^\pi_F(q)\,
\Delta^\rho_{\beta\la}(q_1)
\bar{u}(p')\left(\gamma_\la\,-\,\frac{\kappa_V}{2M} \sigma_{\la\alpha} q_{1\,\alpha}
\right)\tau^a u(p)\,,  \nonumber \\
M^{an,\,a}_{\mu\nu}(2)&\,=\,&-i\frac{g^2_\rho}{3}\,
\vep_{\eta \nu \beta \mu}\,k_\eta\,\Delta^\rho_{\beta\la}(q_1)
\bar{u}(p')\left(\gamma_\la\,-\,\frac{\kappa_V}{2M} \sigma_{\la\alpha} q_{1\,\alpha}
\right)\tau^a u(p)\,,  \label{RMCAan}
\eea
and they correspond to the processes presented in Figs.\,\ref{figRMCAA}(a) and \,\ref{figRMCAA}(b). Together 
with the radiative pion absorption amplitude of Fig.\,\ref{figRMCAA}(c)
\be
M^{an,\,a}_{\pi,\,\nu}\,=\,-\frac{g_\rho}{3f_\pi}\,
\vep_{\eta \nu \beta \sigma}\,k_\eta q_\sigma
\Delta^\rho_{\beta\la}(q_1)
\bar{u}(p')\left(\gamma_\la\,-\,\frac{\kappa_V}{2M} \sigma_{\la\alpha} q_{1\,\alpha}
\right)\tau^a u(p)\,, \label{Manpi}
\ee
the amplitudes Eq.\,(\ref{RMCAan}) satisfy PCAC:
\be
q_\mu\left[M^{an,\,a}_{\mu\nu}(1)\,+\,M^{an,\,a}_{\mu\nu}(2)\right]\,=\,
i f_\pi m^2_\pi \Delta^\pi_F(q)\,M^{an,\,a}_{\pi,\,\nu}\,.  \label{ANPCAC}
\ee

The contribution from the anomalous amplitudes Eq.\,(\ref{RMCAan}) to the total hadron radiative 
amplitude Eq.\,(\ref{Ta}) for the reaction Eq.\,(\ref{RMCp}) is given as
\be
T^a_{an}\,=\,\frac{eG}{\sqrt{2}}l_\mu(0)\varepsilon_\nu\,\left(\tilde{c}_7+\tilde{c}_9\right)
\left[M^{an,\,a}_{\mu\nu}(1)\,+\,M^{an,\,a}_{\mu\nu}(2)\right]\,.  \label{Taan}
\ee
In Ref.\,\cite{FLMS}, a contribution arising from the Wess--Zumino--Witten anomalous
Lagrangian was estimated. Graphically, it corresponds to our Fig.\,\ref{figRMCAA}(b) with the pion instead of
the rho meson and with the vector interaction ${\hat {\cal V}}^a_\mu$ instead of the axial one.
The associated amplitude depends on the momentum transfer $q^L$ and
therefore, it does not possess the enhancement factor $\approx\,3$ for large
photon momentums. For illustration, we present the contribution to one of the form factors
\be
\Delta g_4'\,=\,-\eta\,\frac{k^2}{8\pi^2 f^2_\pi}\frac{\lambda k + y \nu}{2m_\mu}\,g^L_P\,,  \label{dg4p}
\ee
all other contributions have a similar structure. It is also seen that these contributions are of
the order ${\cal O}(1/M^3)$, because $8\pi^2 f^2_\pi\,\sim\,M^2$.

In our case, it is the amplitude related to the graph Fig.\,\ref{figRMCAA}(a), 
which is $q^N$--dependent. We present from the calculated contributions to $g_i$ arising from the 
amplitudes Eq.\,(\ref{RMCAan}) only the one for the form factor $g_2$, the other ones are suppressed
by one order in $1/M$ 
\be
\Delta g_2\,=\,-\,\frac{g^2_\rho}{3g_A}\left(\frac{2M}{m_\rho}\right)^2(1+\kappa_V)\eta\,\frac{k}{2M}\,
\frac{\vec{s}^{\,2}}{4M^2}\,g^N_P\,
 +\,\frac{g^2_\rho}{3}\left(\frac{2M}{m_\rho}\right)^2(1+\kappa_V)\eta\,
\frac{\vec{k}\cdot\vec{s}}{4M^2}\,.  \label{dgisan}
\ee
For comparison, we keep also the $g^N_P$--dependent contribution. Using the KSFR relation 
$2f^2_\pi g^2_\rho= m^2_\rho$, we can rewrite this contribution to the form
\begin{displaymath}
-\frac{1+\kappa_V}{g_A}\,\eta\,\frac{k}{M}\frac{\vec s^{\,2}}{12f^2_\pi}\,g^N_P\,.
\end{displaymath}
Taking into account that $12 f^2_\pi\,\approx\,M^2/10$ we can see that the $g^N_P$--dependent contribution
is larger than the $g^L_P$--dependent one for large values of $k$ by a factor $\approx\,20$ . However,
it is not enough to influence the photon  spectrum, because of an additional factor
$(\tilde{c}_7+\tilde{c}_9)$ in the amplitude Eq.\,(\ref{Taan}) (see below).
It is seen from Eq.\,(\ref{dgisan}) that the first term on the right--hand side
is suppressed in comparison with the second one 
arising from the contact amplitude $M^{an,\,a}_{\mu\nu}(2)$. As we shall see later, the second term
contributes to the triplet capture rate by an amount of $\approx\,-0.2\%$.

Let us note that the sum of the vector RMC amplitudes arising from the anomalous Lagrangian Eq.\,(\ref{Lan}) 
is zero with a good accuracy.

One can obtain more general result for $\Delta g_2$, Eq.\,(\ref{dgisan}), by a change
\begin{displaymath}
1/3\,\rightarrow\,g_{\rho 1}/g_{\omega 1}\,,\quad \kappa_V\,\rightarrow\,g_{\rho 2}/g_{\rho 1}\,,
\end{displaymath}
which coresponds to the model used in Refs.\,\cite{DMW1,DMW2} for describing the pion photoproduction
amplitude in the t channel. In this model, the $\pi \rho \gamma$ and $\pi \omega \gamma$ amplitudes are 
effectively the same as those obtained from our anomalous Lagrangian Eq.\,(\ref{Lan}) and
the $\rho NN$ and $\omega NN$ vertices contain four constants $g_{\rho 1}$, $g_{\rho 2}$, $g_{\omega 1}$
and $g_{\omega 2}$, which are the free parameters, obtained together with other free parameters
of the model from a fit to the data.

Compared with the sets of the form factors given in Eqs.\,(\ref{dgis1}) and (\ref{dgis2}), 
the $g_i$'s, Eq.\,(\ref{dgisan}),
are even larger. However, due to the values  of ${\tilde{c}}_i$, Eq.\,(\ref{cti}), 
the factor $(\tilde{c}_7+\tilde{c}_9)\,\approx\,1.8\times 10^{-2}$
makes the contribution from the amplitude $T^a_{an}$, Eq.\,(\ref{Taan}), small. 

\section{Results}
\label{CH3}
 
Using the Hamiltonian $H^{\,(0)}_{\,eff}$, Eq.\,(\ref{H0eff}), and the sets of 
the form factors $g_i$, Eqs.\,(\ref{gis0}), (\ref{dgis1}),(\ref{dgis2}), and 
(\ref{dgisd}), we have calculated the capture rates and the photon energy 
spectra for the RMC in a muon-hydrogen system described by a spin density 
matrix $\rho_\xi$ ($\xi=s,t$) \bea 
\frac{d\Lambda_\xi}{dk}&\,=\,&\frac{1}{4\pi^3}\left(\alpha^2 
G_F\,cos\,\theta_c\,m/m_\mu\right)^2 m M_n k \int^{+1}_{-1}\,dy 
\frac{\nu^2_0}{W+k(y-1)}\,Tr\left\{\left(1-{\vec \sigma} \cdot \hat \nu\right) 
\right. \nonumber  \\ & & \left. \times 
H^{(0)}_{eff}\,\rho_\xi\,\left[H^{(0)}_{eff}\right]^+\right\}\,. \label{spec} 
\eea Here $\alpha$ is the fine structure constant, $G_F$ is the Fermi constant, 
$cos\,\theta_c$ is the Cabibbo angle, $m$ is the reduced mass of the 
$\mu\,-\,p$ system, the neutrino energy is determined by the energy 
conservation \bea 
\nu_0\,&=&\,\frac{W}{W+k(y-1)}(k_{max}-k)\,\approx\,\left[\,1\,+\,\frac{k}{M_p}(1-y)\,+\,\frac{k}{M_p^2}(y-1) 
\right.\nonumber \\ 
&& \left.\,\times\,\left(m_\mu+k(y-1)\right)\right]\left(k_{max}-k\right)\,,   \label{nu0} 
\eea
where the maximum photon energy is given as 
\be
k_{max}\,=\,\frac{W^2-M^2_n}{2W}\,,\quad
W\,=\,M_p+m_\mu\,,   \label{km}
\ee
and $M_{p(n)}$ is the proton (neutron) mass. The singlet and triplet spin density matrices are \cite{Ba,Sh}
\be
\rho_s\,=\,\frac{1}{4}(1\,-\,\vec \sigma \cdot \vec \sigma_l)\,,\quad 
\rho_t\,=\,\frac{1}{4}(1\,+\,\frac{1}{3}\vec \sigma \cdot \vec \sigma_l)\,.   \label{rhost} 
\ee
We have also calculated the capture rates and spectra for the ortho- and paramolecular p$\mu$p states
and for the mixture of muonic states relevant to the TRIUMF experiment \cite{TRIUMF1,TRIUMF2}.
The ortho  ($\Lambda_o$)- and paramolecular ($\Lambda_p$) capture rates are given in terms 
of $\Lambda_s$ and $\Lambda_t$ as \cite{Ba}
\be
\Lambda_o\,=\,0.756\,\Lambda_s\,+\,0.253\,\Lambda_t\,, \quad 
\Lambda_p\,=\,0.286\,\Lambda_s\,+\,0.857\,\Lambda_t\,,  \label{Lot}
\ee
and the capture rate $\Lambda_T$, relevant to the TRIUMF experiment \cite{TRIUMF1,TRIUMF2}, is
\be
\Lambda_T\,=\,0.061\,\Lambda_s\,+\,0.854\,\Lambda_o\,+\,0.085\,\Lambda_p\,.   \label{LT}
\ee
Now we present numerical results for the capture rates.

\subsection{Capture rates}
\label{CH3.1}

Here we present the results for the capture rates calculated 
in  various models. If not stated otherwise, we use the monopole form factors and we put $x=1$ in Eq.(\ref{gPLN}). 
 
A) We first discuss the results obtained in the model with the $\Delta$ isobar kept on--shell.
We give the singlet and triplet capture rates in a more detail, in order to see 
explicitly various contributions, 
\bea
\Ls\,\times\,10^3\,&=&\,0.40(0)\,+\,1.65(-1)\,+\,1.29(-2)\,+\,0.11(hp)\,-\,0.07(al)\,
                          +\,0.05(\Delta)\,  \nonumber  \\
                   &=&\,3.43\,(3.51)\,\sm1\,,   \label{Ls0}   \\
\Lt\,\times\,10^3\,&=&\,43.7(0)\,+\,53.1(-1)\,+\,3.7(-2)\,+\,0.9(hp)\,-\,0.2\,(al)\,
                          +\,2.2(\Delta)\,   \nonumber  \\
                   &=&\,103.0\,(103.4)\,\sm1\,. \label{Lt0}
\eea 
Here on the right--hand side of Eqs.\,(\ref{Ls0}) and (\ref{Lt0}), the number n 
(n=0,-1,-2) in the brackets means the order of the contribution in ${\cal 
O}(1/M^n)$ and hp (al) and $\Delta$ mean the contributions from the hard pion 
form factors Eq.\,(\ref{dgis2}) (from the form factors Eq.\,(\ref{dgisan})) and from the 
form factors Eq.\,(\ref{dgisd}) due to the $\Delta$ isobar excitation processes, respectively. 
These were calculated using the parameters 
\be
\frac{f^2_{\pi N \Delta}}{4\pi}\,=\,0.371\,,\quad G_1\,=\,2.525\,,\quad Y\,=\,Z=\,-0.5\,. \label{DC}
\ee
The choice of the parameters $Y$ and $Z$ is such that only the terms proportional to $\delta_{\mu \nu}$
in Eq.\,(\ref{Th}) contribute (the $\Delta$ is on--shell). The contribution of the $\Delta$ excitation 
to $\Lt$ is $\approx\,2\%$, which is in agreement with BHM. The numbers in the brackets on the
right--hand side of Eqs.\,(\ref{Ls0}) and (\ref{Lt0}) are obtained using Eq.\,(\ref{gPLN})
for $g^{L(N)}_P$ without the correction ${\tilde g}^{L(N)}_P$ included.

For the other capture rates we have
\be
\Lambda_o\,=\,28.7\tm3\sm1\,,\quad \Lambda_p\,=\,89.3\tm3\sm1\,,\quad
\Lambda_T\,=\,32.3\tm3\sm1\,. \label{LopT0}
\ee

For the dipole form factors, analogously with Eqs.\,(\ref{Ls0}) and (\ref{Lt0}) we obtain
\be
\Ls\,=\,3.28\,(3.50)\tm3\sm1\,,\quad \Lt\,=\,101.5\,(102.5)\tm3\sm1\,.  \label{Lst0d}
\ee

Let us compare our results with available calculations. The singlet and triplet capture rates were
calculated earlier by Opat \cite{O}. He obtained $\Ls=4.96\tm3$s$^{-1}$ and $\Lt=90.0\tm3$s$^{-1}$.

Very recent calculations \cite{BHM} yield $\Ls=(2.90-3.10)\tm3$s$^{-1}$ and 
$\Lt=(112-114)\tm3$s$^{-1}$. Having in mind that $\Ls$ results as the difference of two large and
almost equal numbers, the agreement  between our value Eq.\,(\ref{Ls0}) of $\Ls$ and the BHM value 
can be considered as satisfactory. However, the difference of $\approx$ 10\% between the 
triplet capture rates is too large. A half of this discrepancy can be understood by checking
the integration over the phase volume. We use for the neutrino momentum Eq.\,(\ref{nu0}) with 
$k_{max}$ from Eq.\,(\ref{km}), while BHM employ for $k_{max}$ equation (4.37), which is in our notations
\be
k_{max}\,=\,m_\mu(1+\frac{m_\mu}{2M_N})(1+\frac{m_\mu}{M_N})^{-1}
\,\approx\,m_\mu(1-\frac{m_\mu}{2M_N}+\frac{m^2_\mu}{2M^2_N})\,,  \label{kmBHM}
\ee
where $M_N=(M_p+M_n)/2$ is the nucleon mass. From Eq.\,(\ref{km}), one obtains
$k_{max}=99.15\,$MeV, while from Eq.\,(\ref{kmBHM}) one has $k_{max}=100.3\,$MeV and
instead of $\Lt=103.0\tm3$s$^{-1}$ one obtains $\Lt=108.0\tm3$s$^{-1}$, which is larger by $\approx\,5\%$.
Using in Eq.\,(\ref{nu0}) for $\nu_0$ the expansion Eq.\,(\ref{kmBHM}) for $k_{max}$, one obtains 
\bea
\nu_0\,&=&\,m_\mu-k\,-\,\frac{m^2_\mu}{2M_N}\,+\,\frac{k}{M_N}(1-y)(m_\mu-k)\,+\,\frac{1}{2M^2_N}
\left[\,m_\mu+k(y-1)\,\right]   \nonumber  \\
 && \,\times\,\left[\,m^2_\mu+2k(m_\mu-k)(y-1)\,\right]\,.  \label{nu0BHM}
\eea
In Ref.\,\cite{BHM}, Eq.\,(4.39) is used for $\nu_0$. It retains terms up to the order 
${\cal O}(1/M_N)$\footnote{There is factor 2 missing in the denominator of the third term 
on the right--hand side of this equation.}, 
which yields  in our case $\Lt=107.2\tm3$s$^{-1}$. But the source of the remaining difference 
of $\approx\,5\%$ between the results for the $\Lt$ is not clear.

B) The results of calculations without the $\Delta$ excitation effect are

\be
\Ls=3.38\tm3\sm1\,,\,\, \Lt=100.8\tm3\sm1\,,\,\,  \Lambda_T=31.6\tm3\sm1\,.   \label{crb0}
\ee
C) We now present the capture rates for the same case as in B), but for the value of the parameter $x=1.5$. 
The rates are
\be
\Ls=8.35\tm3\,\sm1\,,\,\, \Lt=116.6\tm3\,\sm1\,,\,\,  \Lambda_T=39.8\tm3\,\sm1\,.   \label{crb1}
\ee
The strong dependence of the capture rates on $g_P$ was already known to Opat \cite{O}.

D) The capture rates, calculated for the same parameters as in A), but $Y=1.75$ and $Z=-0.8$, are
\be
\Ls=3.28\tm3\,\sm1\,,\,\, \Lt=106.0\tm3\,\sm1\,,\,\, \Lambda_T=33.0\tm3\,\sm1\,.   \label{crd3}
\ee
In this case, the effect of the $\Delta$ excitation is for the values of $Y$ and $Z$, allowed by the inequalities of
Eq.\,(\ref{ZYG1}), maximal and it is $\approx\,5\%$ for $\Lt$ and $\approx\,4\%$ for $\Lambda_T$.

E) The capture rates, calculated as in D, but $Z=-1.95$,
\be
\Ls=5.93\tm3\,\sm1\,,\,\, \Lt=116.7\tm3\,\sm1\,,\,\, \Lambda_T=38.1\tm3\,\sm1\,.   \label{crd2}
\ee
These capture rates are close to those calculated in the case C) for $x=1.5$. In order to achieve
the same enhancement in the rates, one should put  $\Delta$ isobar more off--shell than
it is needed in pion photoproduction processes. On the other hand, the values of the off--shell
parameters $X,\,Y,\,Z$ depend strongly on whether the pion production amplitude is unitarized or not
and on the method of unitarization. Let us note that our choice of the parameters of the model
is not optimal. In order to extract an optimal set of these parameters from the data, one should use 
a minimization procedure.

Similar enhancement in the capture rates as in the case E) can be achieved by taking 
$f^2_{\pi N \Delta}/4\pi\,\approx\,20$ or $G_1\,\approx\,20$, which is an amplification of $\approx\,7$
in the $\pi N \Delta$ coupling and of $\approx\,8$ in the constant $G_1$, which is much more than
the enhancement factor of $\approx\,2.5$ for the parameter $Z$.

\subsection{Photon spectra}
\label{CH3.2}

\begin{figure}[htb]
\centerline{
\epsfig{file=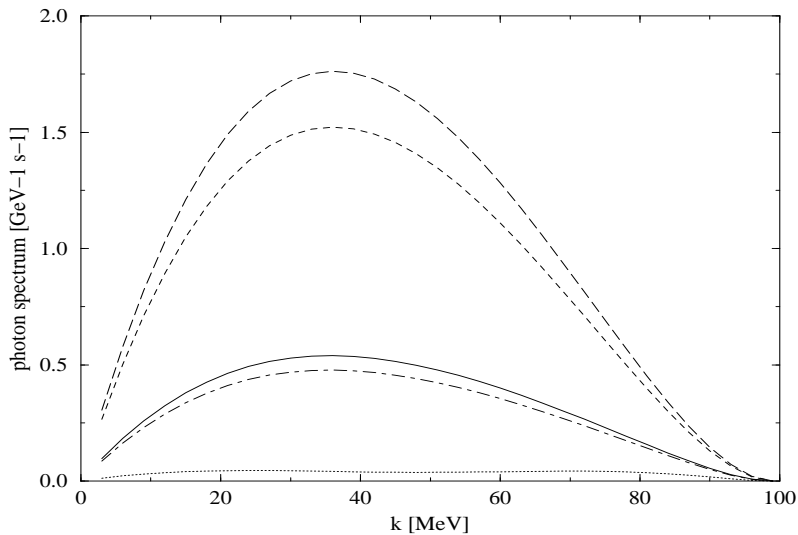,width=7cm,height=10.5cm,angle=270}
}
\vskip 0.4cm
\caption[99]{
Photon spectra calculated in the model A of the previous section, the $\Delta$ isobar excitation effect
is included; the dotted, long--dashed, dot--dashed and dashed  curves correspond,
respectively, to the singlet, triplet, ortho- and para $p\mu p$ molecule spin combinations,
the solid curve correspond to the mixture of muonic states relevant to the TRIUMF experiment
\cite{TRIUMF1,TRIUMF2}. 
}
\label{figsp1}
\end{figure}

Photon spectra, corresponding to the capture rates calculated in the model A
of the previous section, are presented in Fig.\,\ref{figsp1}. The $\Delta$ isobar is included,
but the choice of the parameters $Y=Z=-0.5$ is such that the isobar is on--shell.
As it is seen from Fig.\,\ref{figsp1}, our spectra are in a close correspondence with those of 
Fig.\,3 of Ref.\,\cite{BF1}. However, our spectrum for the triplet state of the $\mu-p$ system
(long--dashed curve) differs from the analogous spectrum of Fig.\,5 \cite{BHM}, as it should be, because
the triplet capture rates differ significantly.

\begin{figure}[htb]
\centerline{
\epsfig{file=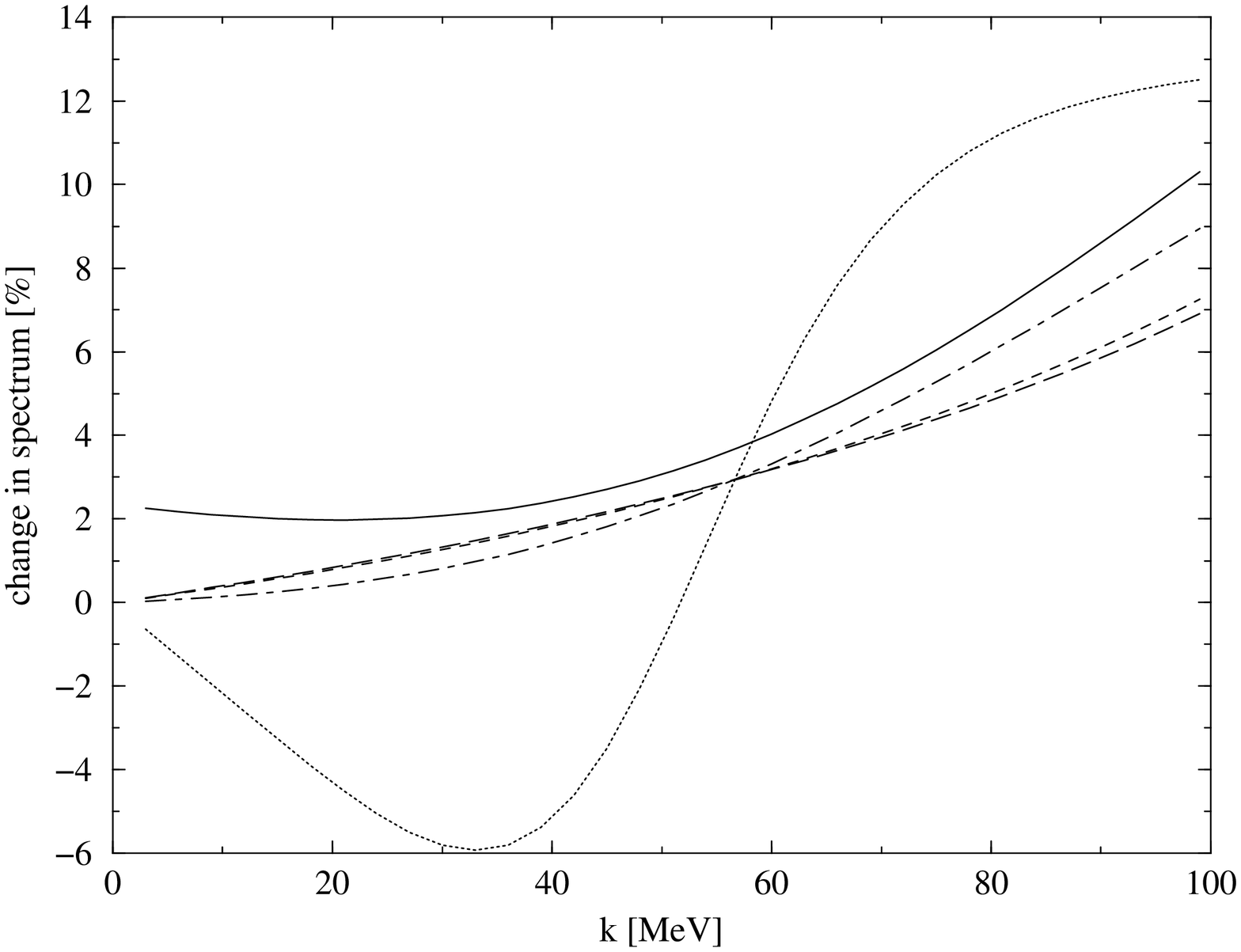,width=7cm,height=10.5cm,angle=270}
}
\vskip 0.4cm
\caption[99]{
Relative difference in \% for the spectra calculated with (model A) and without (model B) 
the $\Delta$ excitation included; the dotted, long--dashed, dot--dashed and dashed  curves correspond,
respectively, to the singlet, triplet, ortho- and para $p\mu p$ molecule spin combinations,
the solid curve correspond to the mixture of muonic states relevant to the TRIUMF experiment
\cite{TRIUMF1,TRIUMF2}.
}
\label{figsp2}
\end{figure}

\begin{figure}[htb]
\centerline{
\epsfig{file=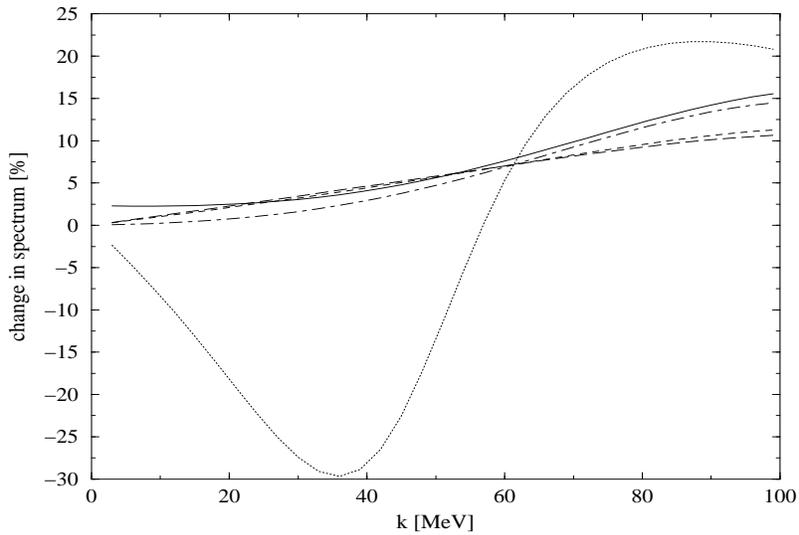,width=7cm,height=10.5cm,angle=270}
}
\vskip 0.4cm
\caption[99]{ 
Relative difference in \% for the spectra calculated with (model D) and without (model B) 
the $\Delta$ excitation included; the dotted, long--dashed, dot--dashed and dashed  curves correspond,
respectively, to the singlet, triplet, ortho- and para $p\mu p$ molecule spin combinations,
the solid curve correspond to the mixture of muonic states relevant to the TRIUMF experiment
\cite{TRIUMF1,TRIUMF2}.
}
\label{figsp3}
\end{figure}

The percentage change in the spectra when the $\Delta$ excitation effect is taken into account
is presented in Fig.\,\ref{figsp2}. The $\Delta$ excitation effect was calculated according to the model A 
of the previous section. The case without this effect corresponds to the model B.
This change in the spectra due to the $\Delta$ was first calculated by Beder and Fearing \cite{BF1}.
Our Fig.\,\ref{figsp2} is in a good agreement with Fig.\,4 of Ref.\,\cite{BF1}, but it differs
from the analogous Fig.\,6 of Ref.\,\cite{BHM}.   

\begin{figure}[htb]
\centerline{
\epsfig{file=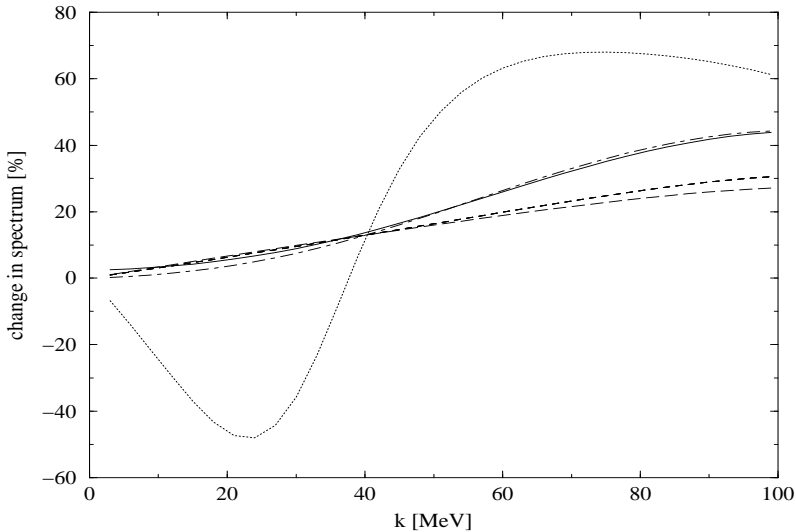,width=7cm,height=10.5cm,angle=270}
}
\vskip 0.4cm
\caption[99]{ 
Relative difference in \% for the spectra calculated with (model E) and without (model B) 
the $\Delta$ excitation included; the dotted, long--dashed, dot--dashed and dashed  curves correspond,
respectively, to the singlet, triplet, ortho- and para $p\mu p$ molecule spin combinations,
the solid curve correspond to the mixture of muonic states relevant to the TRIUMF experiment
\cite{TRIUMF1,TRIUMF2}.
}
\label{figsp4}
\end{figure}

Similar calculations are presented in Fig.\,\ref{figsp3} and Fig.\,\ref{figsp4} 
using instead of the model A the spectra of the model D and E, respectively. As it is seen, 
by putting the $\Delta$ isobar off--shell, 
both the singlet and triplet spectra are changed sensibly.

\begin{figure}[htb]
\centerline{
\epsfig{file=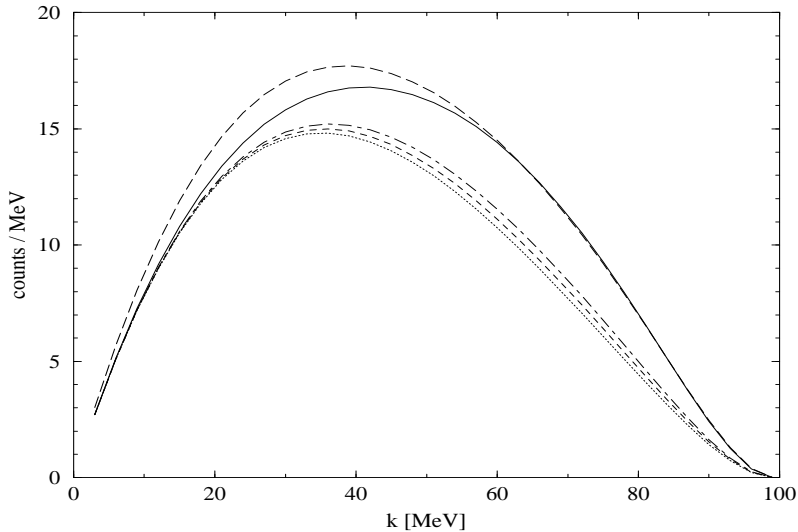,width=7cm,height=10.5cm,angle=270}
}
\vskip 0.4cm
\caption[99]{ Influence of the $\Delta$ isobar parameters on the photon spectra 
corresponding to the mixture of muonic states in TRIUMF experiment
\cite{TRIUMF1,TRIUMF2}. For the explanation of the curves see text.
}
\label{figsp5}
\end{figure}

In Fig.\,\ref{figsp5}, we show how the photon spectra, relevant to the mixture of the
muonic states for the TRIUMF experiment, depend on the parameters
of our model. The dotted curve corresponds to the model B (no $\Delta$, x=1), the dashed
curve is the photon spectrum for the case A ($\Delta$ on--shell, $Y=Z=-0.5$), the dot--dashed curve is calculated
using the model D: $\Delta$ is off--shell, the parameters $Y=1.75,\,Z=-0.8$ are at the boundary of the
region Eq.\,(\ref{ZYG1}) allowed by the pion photoproduction data \cite{DMW1,DMW2,BDM,DPC}. In this case, 
about two times more discrepancy is explained in comparison with the dashed curve. The dependence
of the photon spectrum on the change in $g_P$ is illustrated by the long--dashed curve, calculated within the model C
(no $\Delta$, x=1.5). We have found that this curve, normalized to 14.5 counts at $k=60\,$MeV, follows closely 
the solid curve of Fig.\,4 \cite{TRIUMF2}.  All other curves in this and in the next figure are 
multiplied by this normalization factor.
The long--dashed curve closely follows  the solid curve from $k\,\approx\,60\,$MeV, 
which corresponds to the model E: $\Delta$ is off--shell, the parameter $Y=1.75$ is
the same as in the case D, the parameter $Z=-1.95$. 

Evidently, this picture is in agreement with the one obtained by comparing the Beder--Fearing model \cite{BF1}
with the experiment \cite{TRIUMF1,TRIUMF2}: the experimental photon spectrum can be satisfactorily described with 
$g_P$ of the form Eq.\,(\ref{gPLN}) only for $x\,\approx\,1.5$ and only a small part of the discrepancy
could be explained by calculating the spectra by including the on--shell $\Delta$ isobar. As it is seen
from the solid curve, putting the $\Delta$ isobar off--shell, one can describe the experimental data quite well.

As we have just found, our model
can explain about two times more of the discrepancy between  the $g^{PCAC}_P$, Eq.\,(\ref{gPCACp}), and
of  $g_P$ extracted from the experiment \cite{TRIUMF1,TRIUMF2}, if one restricts oneself
to the set of the parameters from Eq.\,(\ref{ZYG1}). They were extracted
in Refs.\,\cite{DMW1,DMW2,BDM,DPC} from the data on pion photoproduction by using unitarized multipoles arising from
the pion photoproduction amplitude. The problem with the unitarity appears in the pion photoproduction,
because of the pion--nucleon interaction in the final state.
It is true that the time reversed of the pion production amplitude is connected with the 
hadron radiative part of our RMC amplitude by the continuity equations Eqs.\,(\ref{PCAC}),(\ref{divMD}),
and (\ref{ANPCAC}). However, we do not need to unitarize our amplitude. 
Therefore it is not clear, how the
inequalities Eq.\,(\ref{ZYG1}) are restrictive for the problem considered here.
It is seen that the model can reproduce reasonably the experimental
photon spectrum \cite{TRIUMF1,TRIUMF2} in the region $k\,\ge\,60\,$MeV for the PCAC value of $g_P$, Eq.\,(\ref{gPCACp}), 
if the values of the parameters $Y$ and $Z$ are taken to be $Y=1.75$ and $Z=-1.95$.

\begin{figure}[htb]
\centerline{
\epsfig{file=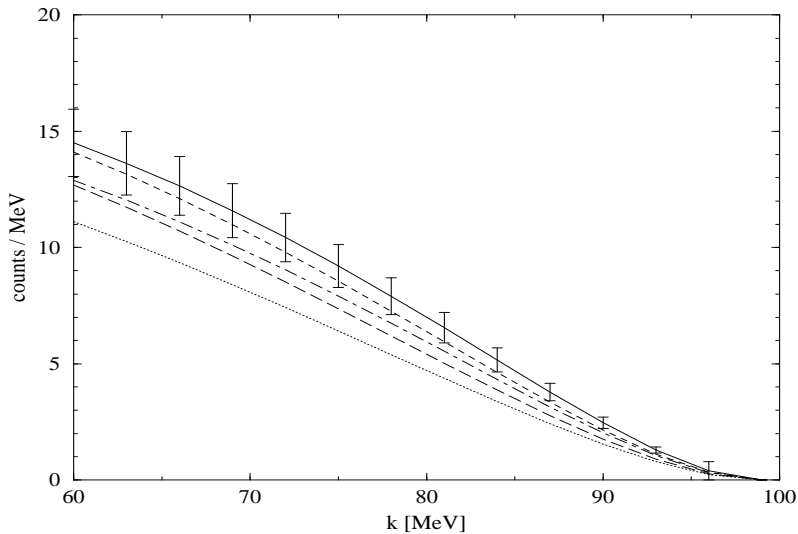,width=7cm,height=10.5cm,angle=270}
}
\vskip 0.4cm
\caption[99]{ 
Dependence of the photon spectra, corresponding to the mixture of muonic states relevant to the TRIUMF experiment
\cite{TRIUMF1,TRIUMF2}, on various parameters of the calculations; the dotted curve is for the reference
and corresponds to the model A (32.3/7.1/196); 
the long--dashed curve is calculated for the model D, but $\xi=0.95$ (36.3/8.1/225); the dot--dashed curve 
is obtained within the model D, but $x=1.2$ (35.9/8.6/239);
the dashed curve-- the model D, but $x=1.2$ and $\xi=0.95$ (39.4/9.3/259); the solid curve corresponds to 
the model C (39.7/9.9/276). For the details see the text.
}
\label{figsp6}
\end{figure}

In Fig.\,\ref{figsp6}, we show the dependence of the calculations on uncertainties in our knowledge
of $g_P$ and of the admixture $\xi$ of the $S=3/2$ orthomolecular $p\mu p$ state.
As discussed recently \cite{AMK1},
admixture of the $S=3/2$ orthomolecular $p\mu p$ state changes the molecular capture rate 
to \cite{W}
\be
\Lambda'_o\,=\,\xi\Lambda_o(1/2)\,+\,(1-\xi)\Lambda_o(3/2)\,,  \label{Lp0}
\ee
where $\Lambda_o(1/2)$ is $\Lambda_o$ of Eq.\,(\ref{Lot}) and $\Lambda_o(3/2)=1.009\,\Lt$. It was found in
\cite{AMK1} that the data on OMC in hydrogen requires $g_P\,\le\,1.2\,g^{PCAC}_P$ or $\xi\,\ge\,0.95$. This restriction
on $g_P$ is in agreement with our Eq.\,(\ref{gPlgOMCHe3}) for the OMC in $^{3}$He. The dependence on the uncertainty
in $g_P$ and on $\xi$ is illustrated by the long--dashed, dot--dashed and dashed curves. The numbers in the brackets
are for the total capture rate, for the partial capture rate in the interval ($60-k_{max})\,$MeV and 
for the capture rate in counts for this
interval, respectively. Otherwise, the unit for the capture rates is $10^{-3}$s$^{-1}$.
For the sake of illustration, we assigned a 10\% error \cite{TRIUMF2} to a set of 'data' 
represented by 14 points of the spectrum C. The number of counts 276, related to this curve is to
be compared with the number of counts 286, which can be read off the histogram presented in Fig.4 \cite{TRIUMF2}.
As it is seen, all curves lay already inside the 2$\sigma$ bound.

\section{Summary}
\label{CH4}

In this paper, we have presented the capture rates and the photon energy spectra for the RMC in hydrogen,
calculated using the effective Hamiltonian $H^{\,(0)}_{\,eff}$, Eq.\,(\ref{H0eff}), where the form factors 
$g_i$ are obtained from the amplitudes
derived from the chiral Lagrangian of the $N \Delta \pi \rho \omega a_1$ system. The non--resonant part
of the Lagrangian contains the normal and anomalous Lagrangians of the $N \pi \rho \omega a_1$
system interacting with the external electromagnetic and weak fields by the associated
one--body currents \cite{IT,STK,KT,STK1,TSK}. In the expansion of the amplitudes in $1/M$, 
we keep all terms up to the order ${\cal O}(1/M^2)$.

For the resonant part of our Lagrangian, we have extended the standardly used model 
\cite{IT,BF1,BHM,KT}
by adopting results of the
model developed by Olsson and Osypowski \cite{OO} and by Davidson, Mukhopadhyay and Wittman \cite{DMW1,DMW2},
allowing the $\Delta$ isobar to be off--shell.  

The calculated triplet capture rate differs by $\approx\,10\%$ from the one derived quite recently within the HBChPT approach
in Ref.\,\cite{BHM}.
About a half of this discrepancy can be understood by the use of an approximate equation for the neutrino momentum
in Ref.\,\cite{BHM}. The origin of the rest of the discrepancy is not clear. 

In the model, restricting the $\Delta$ isobar on--shell, our spectra are close to those obtained earlier by
Beder and Fearing \cite{BF1}. 
Our full model, that includes off--shell $\Delta$ isobar,
can explain two times more of the discrepancy between the PCAC value of $g_P$, Eq.\,(\ref{gPCACp}), and
of $g_P$ extracted from the experiment \cite{TRIUMF1,TRIUMF2},
if one restricts oneself to
the values of the parameters of the model extracted from the data on the pion photoproduction off the nucleon. 
Let us note that taking into account existing uncertainty in $g_P^{PCAC}$ extracted from  
the OMC in the hydrogen and $^{3}$He and in
the parameter $\xi$, regulating the admixture of the $S=3/2$ orthomolecular $p\mu p$ state, one finds
that this model is, actually, in reasonable agreement with the data \cite{TRIUMF1,TRIUMF2}.
 
Moreover, if one is allowed to vary the parameters of the model independently, the experimental photon 
spectrum can be described for the induced pseudoscalar 
form factor $g_P$ of Eq.\,(\ref{gPLN}) without any scaling ($x=1$). It would be difficult to find
any physics behind the scaling of $g_P$, which would mean violation of PCAC. On the other hand, 
the part of our full model, containing the electroweak interaction of the off-shell $\Delta$ isobar, 
is widely used to describe successfully the pion photoproduction data,
which provides a strong basis for confidence in our results.
For the reaction of the RMC in the hydrogen, Eq.\,(\ref{OMCp}), it is the only model known so far, providing enough enhancement 
in the high energy region of the photon spectrum.

In conclusion we note that the reactions of the RMC in the hydrogen and $^{3}$He are at present the only
available effective tool for the study of the form factor $g_P$ as a function of the momentum transfer.
Therefore, more efforts, both theoretical and experimental, are highly desirable.

\section*{Acknowledgments}

This work is supported by Grant No. GA \v{C}R 202/00/1669. The research of F.~C.~K.~is
supported in part by NSERCC. A part of this work was done during stays of E.~T.~at the
Theoretical Physics Institute of the University of Alberta. He thanks Professor F.~C.~Khanna for
the warm hospitality. The  help with computations by R.~Teshima is acknowledged. We
thank Professor R.~Davidson for the correspondence and Dr.~O.~Dragoun for discussions.

\end{document}